\documentclass[12pt]{iopart}

\usepackage[english]{babel}

\usepackage{iopams}

\usepackage{amsfonts}
\usepackage{amsthm}
\usepackage{amssymb}

\usepackage{varioref}

\usepackage[
            pdfpagelabels,hypertexnames=true,
            plainpages=false,
            naturalnames=false]{hyperref}

\usepackage{color}
\definecolor{darkblue}{rgb}{0,0.1,0.5}
\hypersetup{colorlinks,
            linkcolor=darkblue,
            anchorcolor=darkblue,
            citecolor=darkblue}

\labelformat{equation}{\textup{(#1)}}

\newtheorem{thm}{Theorem}[section]

\newtheorem{lem}[thm]{Lemma}
\theoremstyle{definition}
\newtheorem{defin}{Definition}[section]
\theoremstyle{remark}
\newtheorem*{remark}{Remark}

\begin{document}

\title{Minimal data at a given point of space for solutions to certain geometric systems}
\author{Andr\'es E. Ace\~na}
\address{Max Planck Institute for Gravitational Physics, Am M\"uhlenberg 1, D-14476 Golm, Germany}
\ead{acena@aei.mpg.de}

\begin{abstract}
We consider a geometrical system of equations for a three dimensional Riemannian manifold. This system of equations has been constructed as to include several physically interesting systems of equations, such as the stationary Einstein vacuum field equations or harmonic maps coupled to gravity in three dimensions. We give a characterization of its solutions in a neighbourhood of a given point through sequences of symmetric trace free tensors (referred to as `null data'). We show that the null data determine a formal expansion of the solution and we obtain necessary and sufficient growth estimates on the null data for the formal expansion to be absolutely convergent in a neighbourhood of the given point. This provides a complete characterization of all the solutions to the given system of equations around that point.
\end{abstract}


\section{Introduction}

In looking for solutions of Einstein field equations, or in analyzing properties of the solutions, it is useful to consider first the static or stationary case. If dimensions higher than four are considered, as in Kaluza-Klein type theories, dimensional reduced versions of the theories are usually also analyzed. This practice is due to the complicated structure of Einstein equations and because the static or stationary solutions provide important information about some properties of the solutions. If one looks at the systems of equations that arise after the dimensional reduction one finds that in general they have the following features:
\begin{itemize}
\item The base manifold is three dimensional.
\item The metric is Riemannian.
\item On the base manifold there are some scalar fields.
\item The metric satisfies an equation whose principal part is given by the Ricci tensor of the metric.
\item The fields satisfy Laplace type equations.
\end{itemize}
There are two features that are central for our analysis. The first is that in a suitable gauge the systems are elliptic. This implies that the solutions are analytic in this gauge. Secondly, in this gauge the systems of equations are diagonal, i.e. for each system the principal symbol has only diagonal elements (the off-diagonal elements vanish) and all the diagonal elements are the same. This implies that the complex null cone (characteristic cone) is the same for all the unknowns in each system. This is very important for the geometrical analysis of the problem.

One of the simplest cases that exhibit the described features is that of the static Einstein vacuum field equations. More complicated examples are given by the stationary Einstein-Maxwell fields in arbitrary dimensions \cite{Ida03} (here stationary means that the dimensionally reduced problem is three dimensional) and static Einstein-Maxwell-dilaton fields \cite{Mars01}. A particular class of systems that has received increasing attention over the years and that also posses the described properties corresponds to harmonic maps coupled to gravity (also called $\sigma$-models) whose base space is a three dimensional manifold or where the spacetime is stationary \cite{Schimming88,Breitenlohner84}. Of course this listing is far from complete, but it clearly shows that many interesting systems of equations share the mentioned properties.

We want to analyze the existence and characterization of solutions to such systems of equations without entering into the details of how such systems of equations arise or what the specific physical considerations that lead to such systems are. Therefore we consider a more general system of equations that includes all the named features. The setting is the following.

Let $N$ be a three dimensional manifold with negative definite metric $h$ and $n$ scalar fields on it, $\phi^\alpha$, $\alpha=1,\dots,n$, that satisfy the following system of equations:
\begin{equation}\label{lapPhi}
 \Delta_h\phi^\alpha=f^\alpha(\phi^\gamma,D_{c}\phi^\gamma),\hspace{1cm}\alpha=1,\dots,n,
\end{equation}
\begin{equation}\label{Riccih}
 R_{ab}[h]=F^\alpha(\phi^\gamma) D_aD_b\phi^\alpha+F^{\alpha\beta}(\phi^\gamma)D_a\phi^\alpha D_b\phi^\beta+f(\phi^\gamma,D_c\phi^\gamma) h_{ab},
\end{equation}
where $D$ is the covariant derivative associated with $h$, $\Delta_h$ is the Laplace operator with respect to $h$ and $R_{ab}[h]$ is the Ricci tensor of $h$.  $F^\alpha(\phi^\gamma)$, $F^{\alpha\beta}(\phi^\gamma)$, $\alpha,\beta=1,\dots,n$, are analytic functions of the scalar fields and $f(\phi^\gamma,D_{c}\phi^\gamma)$, $f^\alpha(\phi^\gamma,D_c\phi^\gamma)$, $\alpha=1,\dots,n$, are analytic functions of the scalar fields and its first derivatives. In order not to burden the text, the span of greek indices is not always stated. We assume that the summation rule for repeated indices holds for both latin and greek indices, the latter being written always in the upper position, as no metric is associated with them. We assume $F^{\alpha\beta}=F^{\beta\alpha}$, so we do not to have to symmetrize on the $a,b$ indices. The functions $f$, $f^\alpha$, $F^\alpha$, $F^{\alpha\beta}$ are scalars if considered as functions on $N$ through their dependence on the fields and their first derivatives. This is necessary for \ref{lapPhi}, \ref{Riccih} to be a geometrically meaningful system of equations. Later on, we will see that the functions $f$, $f^\alpha$, $F^\alpha$ and $F^{\alpha\beta}$ are not independent but have to satisfy some relations. We refer to \ref{lapPhi}, \ref{Riccih} as the field equations.

Now we can state clearly the objective of this work: to give a complete characterization of the solutions to the field equations above in a neighbourhood of a given point.

So, let us take a point $q$ in $N$. As said before, in a suitable gauge the system of equations is elliptic, therefore the solution, if it exists, is analytic. This suggest prescribing as data the coefficients of the Taylor expansion of the fields at $q$. This is not possible, as the Taylor expansion coefficients are not independent of each other. Instead we consider the following $n$ sequences of symmetric trace-free tensors at $q$,
\begin{equation}\label{nullDataGeneral}
 {\cal D}^\alpha=\{\phi^\alpha(q),D_{a_1}\phi^\alpha(q),{\cal C}(D_{a_2}D_{a_1}\phi^\alpha)(q),{\cal C}(D_{a_3}D_{a_2}D_{a_1}\phi^\alpha)(q),\dots\},
\end{equation}
where ${\cal C}$ means taking the symmetric trace-free part of the tensor to which it is applied. These sequences are called the {\it null data} for our system of equations. It turns out that the null data is indeed a minimal set of data for the field equations, that is, it determines a formal expansion of the solution and its components are independent of each other. Our purpose can then be stated as to derive necessary and sufficient conditions for the null data to determine apart from gauge conditions (unique) real analytic solutions of \ref{lapPhi} and \ref{Riccih}. This will show that the null data do indeed provide the searched characterization.

Prescribing a minimal set of data at a point as characterization for the solutions to a gauge-elliptic problem is certainly different from posing a standard boundary value problem. A particular advantage of our approach is that the data has a geometrical meaning and does not depend on the choice of an arbitrary hypersurface or coordinate system. This makes the characterization intrinsic to the geometry of the solution.

Another advantage arises if there is a geometrically distinguished point. Our approach then allows a complete control and analysis of the solution at the given point. The existence of a geometrically distinguished point happens in the case of asymptotically flat static or stationary spacetimes (cf. Friedrich \cite{Friedrich07} and Ace\~na \cite{Acena09}), where there is a point in the manifold that represents infinity. Unfortunately those cases are not included in our present treatment because besides the metric there are further tensor fields among the unknowns, while we only consider scalar fields. An extension of the present result to include tensors in the unknowns is possible although not entirely straightforward, as one has possibly to take into account integrability conditions that may arise for the system of equations to be consistent. We do not want to get involved in such a discussion here.

It is convenient to express the tensors in ${\cal D}^\alpha$ in terms of an $h$-orthonormal frame $c_{\bf a}$, ${\bf a}=1,2,3,$ centered at $q$. Denoting by $D_{\bf a}$ the covariant derivative in the direction of $c_{\bf a}$,
\begin{equation}\label{nullData}
 {\cal D}^{\alpha*}=\{\phi^\alpha(q),D_{\bf a_1}\phi^\alpha(q),{\cal C}(D_{\bf a_2}D_{\bf a_1}\phi^\alpha)(q),{\cal C}(D_{\bf a_3}D_{\bf a_2}D_{\bf a_1}\phi^\alpha)(q),\dots\}.
\end{equation}
These tensors will be called the {\it null data in the frame $c_{\bf a}$}, and are defined uniquely up to rigid rotations in ${\mathbb R}^3$.

If the metric $h$ and the potentials $\phi^\alpha$ exist, then they are real analytic near $q$ and one has Cauchy estimates on the derivatives of the potentials. The Cauchy estimates imply that there exist positive constants $M$, $r$, such that the components of the null data satisfy
\begin{equation}\label{nullDataEstimates}
|{\cal C}(D_{{\bf a}_p}...D_{{\bf a}_1}\phi^\alpha)(q)|\leq \frac{Mp!}{r^p},\hspace{1cm}p\geq 0,\hspace{1cm}{\bf a}_p,\dots,{\bf a}_1=1,2,3.
\end{equation}
Our main result corresponds to the statement that these estimates are not only necessary but also sufficient to have an analytic solution, and is presented in the following theorem.

\begin{thm}\label{mainThm}

Suppose
\begin{equation}\label{abstractNullData}
\hat{{\cal D}}^\alpha=\{\psi^\alpha_{{\bf a}_1},\psi^\alpha_{{\bf a}_2{\bf a}_1},\psi^\alpha_{{\bf a}_3{\bf a}_2{\bf a}_1},...\},\hspace{1cm}\alpha=1,\dots,n,
\end{equation}
are $n$ infinite sequences of symmetric, trace free tensors given in an orthonormal frame at the origin of a 3-dimensional Euclidean space. If there exist positive constants $M$, $r$ such that the components of these tensors satisfy the estimates
\begin{equation*}
|\psi^\alpha_{{\bf a}_p...{\bf a}_1}|\leq \frac{Mp!}{r^p},\hspace{1cm}p\geq 0,\hspace{1cm}{\bf a}_p,...,{\bf a}_1=1,2,3,\hspace{1cm}\alpha=1,\dots,n,
\end{equation*}
then there exists an analytic solution $h$, $\phi^\alpha$, $\alpha=1,\dots,n$, of the field equations near $q$, unique up to isometries, so that the null data implied by it in a suitable frame $c_{\bf a}$ as described above satisfy
\begin{equation*}
\hspace{-1cm}{\cal C}(D_{{\bf a}_p}...D_{{\bf a}_1}\phi^\alpha)(q)=\psi^\alpha_{{\bf a}_p...{\bf a}_1},\hspace{0.5cm}p\geq 0,\hspace{0.5cm}{\bf a}_p,...,{\bf a}_1=1,2,3,\hspace{0.5cm}\alpha=1,\dots,n.
\end{equation*}
\end{thm}

The sequences \ref{abstractNullData}, not necessarily satisfying any estimates, will be referred to as \emph{abstract null data}. As the type of estimates imposed here on the abstract null data does not depend on the orthonormal frame in which they are given, and since these estimates are necessary as well as sufficient, then all possible solutions of \ref{lapPhi}, \ref{Riccih} are characterized by the null data.

In the context of three dimensional Riemannian spaces (cf. Penrose and Rindler \cite{PenroseRindler87} for the four dimensional Lorentzian case) the null data was first introduced by Friedrich \cite{Friedrich07} as a way to characterize static asymptotically flat solutions to the vacuum Einstein's field equations. They were also used by Ace\~na \cite{Acena09} for the stationary asymptotically flat vacuum case. The techniques that we use to prove the result of the present work are similar to those introduced by Friedrich and used by Ace\~na. Therefore we will not present the procedure in full detail, but we will state the important steps and the features that are distinctive for the case that we are treating here.

\section{The exact sets of equations argument}\label{exactSets}

We have defined the null data ${\cal D}^\alpha$ \ref{nullDataGeneral} and we want to use them to characterize solutions to the field equations. Therefore an important first step is to show that the null data can actually be used to construct formal solutions to the field equations. For this we construct expansions of the fields in normal coordinates.

We assume from now on $N$ to be small enough to coincide with a convex $h$-normal neighbourhood of $q$. Let $c_{\bf a}$, ${\bf a}=1,2,3$, be an $h$-orthonormal frame field on $N$ which is parallelly transported along the $h$-geodesics through $q$ and let $x^a$ denote normal coordinates centered at $q$ so that $c^b\,_{\bf a}\equiv \langle dx^b,c_{\bf a}\rangle=\delta^b\,_{\bf a}$ at $q$. We refer to such a frame as a \emph{normal frame centered at} $q$. Its dual frame will be denoted by $\chi^{\bf c}=\chi^{\bf c}\,_b dx^b$. In the following all tensor fields, except the frame field $c_{\bf a}$ and the coframe field $\chi^{\bf c}$, will be expressed in terms of this frame field, so that the metric is given by $h_{\bf ab}\equiv h(c_{\bf a},c_{\bf b})=-\delta_{\bf ab}$. With $D_{\bf a}\equiv D_{c_{\bf a}}$ denoting the covariant derivative in the $c_{\bf a}$ direction, the connection coefficients with respect to $c_{\bf a}$ are defined by $D_{\bf a}c_{\bf c}=\Gamma_{\bf a}\,^{\bf b}\,_{\bf c}c_{\bf b}$.

An analytic tensor field $T_{{\bf a}_1...{\bf a}_k}$ on $N$ has in the normal coordinates $x^a$ a \emph{normal expansion} at $q$, which can be written
\begin{equation}\label{normalExpansion}
T_{{\bf a}_1...{\bf a}_k}(x)=\sum_{p\geq 0}\frac{1}{p!}x^{b_p}...x^{b_1}D_{{\bf b}_p}...D_{{\bf b}_1}T_{{\bf a}_1...{\bf a}_k}(q),
\end{equation}
where we assume from now on that the summation convention does not distinguish between bold face and other indices.

Since $h_{\bf ab}=-\delta_{\bf ab}$, it remains to be seen how to obtain normal expansions for the $\phi^\alpha$'s using the field equations and the null data. That is, we need to see how to obtain
\begin{equation*}
 D_{{\bf a}_p}...D_{{\bf a}_1}\phi^\alpha(q),\hspace{1cm}p\geq 0,\hspace{1cm}{\bf a}_p,\dots,{\bf a}_1=1,2,3.
\end{equation*}
The algebra necessary for doing this simplifies considerably in the space-spinor formalism. How to do the transition is explained in \cite{Friedrich07}. Here we recall a few important properties.
\begin{itemize}
\item{A space spinor field $T_{A_1B_1...A_pB_p}=T_{(A_1B_1)...(A_pB_p)}$ arises from a real tensor field $T_{{\bf a}_1...{\bf a}_p}$ if and only if
\begin{equation}\label{realityCond}
T_{A_1B_1...A_pB_p}=(-1)^p\tau_{A_1}\,^{A_1'}..\tau_{B_p}\,^{B_p'}\bar{T}_{A'_1B'_1...A'_pB'_p},
\end{equation}
where $\tau^{AA'}=\epsilon_0\,^A\epsilon_0\,^{A'}+\epsilon_1\,^A\epsilon_1\,^{A'}$. $\epsilon_{AB}$ is the constant $\epsilon$-spinor, which satisfies $\epsilon_{AB}=-\epsilon_{BA}$, $\epsilon_{01}=1$ and it is used to move indices according to the rules $\iota_B=\iota^A\epsilon_{AB}$, $\iota^A=\epsilon^{AB}\iota_B$.}
\item{Any spinor field $T_{A...H}$ admits a decomposition into products of totally symmetric spinor fields and $\epsilon$-spinors which can be written schematically in the form
\begin{equation}\label{symmetricCont}
T_{A...H}=T_{(A...H)}+\sum \epsilon's\,\times\,\mbox{\emph{symmetrized contractions of }}T.
\end{equation}}
\item{The operation of taking the symmetric trace-free part of a tensor translates into taking the totally symmetric part of the corresponding spinor. So the null data translates into $n$ sequences of totally symmetric spinors.}
\item{We also have a complex frame field $c_{AB}$ (related to $c_{\bf{a}}$), such that $h(c_{AB},c_{CD})=h_{ABCD}\equiv -\epsilon_{A(C}\epsilon_{D)B}$, and its dual 1-form field $\chi^{AB}$ (related to $\chi^{\bf{a}}$). The covariant derivative of a spinor field $\iota^C$ in the direction of $c_{AB}$ is given by
\begin{equation*}
D_{AB}\iota^C=c_{AB}(\iota^C)+\Gamma_{AB}\,^C\,_D\iota^D,
\end{equation*}
where $\Gamma_{ABCD}=\Gamma_{(AB)(CD)}$ are the spinor connection coefficients.}
\item{The commutator of derivatives are given in terms of the curvature spinor,
\begin{equation}\label{spinorCommutator}
(D_{CD}D_{EF}-D_{EF}D_{CD})\iota^A=r^A\,_{BCDEF}\iota^B,
\end{equation}
where
\begin{eqnarray}\label{spinorRicci}
\hspace{-2cm}r_{ABCDEF} & = & \frac{1}{2}\Bigg[\left(s_{ABCE}-\frac{1}{6}rh_{ABCE}\right)\epsilon_{DF}+\left(s_{ABDF}-\frac{1}{6}rh_{ABDF}\right)\epsilon_{CE}\Bigg],
\end{eqnarray}
being $r$ the Ricci scalar of $h$ and $s_{ABCD}=s_{(ABCD)}$ the trace free part of the Ricci tensor of $h$. In tensor notation the decomposition of the Ricci tensor reads $R_{ab}[h]=s_{ab}[h]+\frac{1}{3}r[h]h_{ab}$.}
\end{itemize}
Equations \ref{lapPhi}, \ref{Riccih} take in the space-spinor formalism the form
\begin{equation}\label{lapPhiSpinor}
\hspace{-1cm} D^P\,_AD_{BP}\phi^\alpha=-\frac{1}{2}\epsilon_{AB}f^\alpha,
\end{equation}
\begin{equation}\label{ricciSpinor}
\hspace{-1cm} r_{ABCDEF}=\frac{1}{2}\Big[\Big(S_{ABCE}-\frac{1}{6}Rh_{ABCE}\Big)\epsilon_{DF}+\Big(S_{ABDF}-\frac{1}{6}Rh_{ABDF}\Big)\epsilon_{CE}\Big],
\end{equation}
where
\begin{equation}\label{spinorS}
 S_{ABCD}=F^\alpha D_{AB}D_{CD}\phi^\alpha+F^{\alpha\beta}D_{AB}\phi^\alpha D_{CD}\phi^\beta+\tilde{f}h_{ABCD},
\end{equation}
\begin{equation}\label{spinorR}
 R=\hat{f},
\end{equation}
being
\begin{eqnarray*}
\tilde{f}=-\frac{1}{3}(F^{\alpha\beta}D_a\phi^\alpha D^a\phi^\beta+F^\alpha f^\alpha),\\
\hat{f}=F^{\alpha\beta}D_a\phi^\alpha D^a\phi^\beta+3f+F^\alpha f^\alpha.
\end{eqnarray*}

Using these equations and the theory of `exact sets of fields' is possible to prove the following result.
\begin{lem}\label{formalExpansion}
 Let there be $n$ given sequences
\begin{equation*}
\hat{{\cal D}}^\alpha=\{\psi^\alpha,\psi^\alpha_{A_1B_1},\psi^\alpha_{A_2B_2A_1B_1},\psi^\alpha_{A_3B_3A_2B_2A_1B_1},...\}
\end{equation*}
of totally symmetric spinors satisfying the reality condition \ref{realityCond}. Assume that there exists a solution $h$, $\phi^\alpha$, to the field equations so that the spinors given by $\hat{{\cal D}}^\alpha$ coincide with the null data ${\cal D}^{\alpha*}$ given by \ref{nullData} in terms of an $h$-orthonormal normal frame centered at $q$, i.e.
\begin{equation*}
\psi^\alpha_{A_pB_p...A_1B_1}=D_{(A_pB_p}...D_{A_1B_1)}\phi^\alpha(q),\hspace{1cm}p\geq 0,
\end{equation*}
Then the coefficients of the normal expansions \ref{normalExpansion} of the fields $\phi^\alpha$, i.e.
\begin{equation*}
 D_{A_pB_p}...D_{A_1B_1}\phi^\alpha(q),\hspace{1cm}p\geq 0,
\end{equation*}
are uniquely determined by the data $\hat{{\cal D}}^\alpha$ and satisfy the reality condition.
\end{lem}
\begin{proof}
The proof is by induction. It holds $\phi^\alpha(q)=\psi^\alpha$, $D_{AB}\phi^\alpha(q)=\psi^\alpha_{AB}$.

To discuss the induction step we assume that the expansion coefficients of $\phi^\alpha$ up to order $p$ are known and start with $D_{A_{p+1}B_{p+1}}...D_{A_1B_1}\phi^\alpha(q)$ and its decomposition in the form \ref{symmetricCont}. By assumption, the totally symmetric part of it is given by $\psi^\alpha_{A_{p+1}B_{p+1}...A_1B_1}$. The other terms in the decomposition contain contractions. Let us consider a general contraction, say $A_i$ contracted with $A_j$. We can commute the operators $D_{A_iB_i}$ and $D_{A_jB_j}$ with other covariant derivatives, generating by \ref{spinorCommutator} and \ref{ricciSpinor} only terms of lower order, until we have
\begin{equation*}
\hspace{-1cm} D_{A_{p+1}B_{p+1}}...D_{A_{i+1}B_{i+1}}D_{A_{i-1}B_{i-1}}...D_{A_{j+1}B_{j+1}}D_{A_{j-1}B_{j-1}}...D_{A_1B_1}D^P\,_{B_{i}}D_{PB_{j}}\phi(q).
\end{equation*}
Equation \ref{lapPhiSpinor} then shows how to express the resulting term by quantities of lower order that are already known.

That the expansion coefficients satisfy the reality condition is a consequence of the formalism and the fact that they are satisfied by the data.
\end{proof}
In order to show the convergence of the formal series determined in the previous lemma we need to impose estimates on the free coefficients given by $\hat{{\cal D}}^\alpha$. For the necessary part we have the following result.
\begin{lem}\label{estimatesNullData}
A necessary condition for the formal series determined in Lemma \ref{formalExpansion} to be absolutely convergent near the origin is that the data given by $\hat{{\cal D}}^\alpha$ satisfy estimates of the type
\begin{equation}\label{estimates-psi}
|\psi^\alpha_{A_pB_p...A_1B_1}|\leq \frac{p!M}{r^p},\hspace{1cm}p\geq 0,
\end{equation}
with some positive constants $M$, $r$.
\end{lem}
The proof of the last lemma is very similar to the respective lemma in \cite{Friedrich07} and is not repeated here.

A maybe more clarifying way of stating the last two lemmas is the following.
\begin{itemize}
\item{If we have two solutions of the field equations and the null data \ref{nullData} of one of the solutions is related to the null data of the other solution by a rigid rotation in $\mathbb{R}^3$, then the two solutions are related by an isomorphism and are therefore geometrically equivalent.}
\item{If we have a solution of the field equations, then its null data \ref{nullData} satisfy estimates of the form \ref{nullDataEstimates}.}
\end{itemize}

\section{The characteristic initial value problem}\label{characteristicProblem}

After showing that the null data determine the solution, one would have to show that the estimates \ref{estimates-psi}, imply Cauchy estimates for the expansion coefficients
\begin{equation*}
|D_{A_pB_p}...D_{A_1B_1}\phi^\alpha(q)|\leq \frac{p!M}{r^p},\hspace{1cm}p\geq 0.
\end{equation*}
This would ensure the convergence of the normal expansion in a neighbourhood of $q$ and the existence of the solution. But deriving estimates on the expansion coefficients from estimates on the null data using the procedure described in the proof of Lemma \ref{formalExpansion} has not been possible. Instead, one can use the intrinsic geometric nature of the problem and the data to formulate the problem as a boundary value problem to which Cauchy-Kowalevskaya type arguments apply. The formalism necessary for this has been developed in \cite{Friedrich07} and used in \cite{Acena09}. Here we present the important facts for the present work, following the notation in \cite{Acena09}. The reader is referred to \cite{Friedrich07,Acena09} for details.

The fields $h$, $\phi^\alpha$ can be extended near $q$ by analyticity into the complex domain and considered as holomorphic fields on a complex analytic manifold $N_c$. Under the analytic extension, and choosing $N_c$ to be a sufficiently small neighbourhood of $q$, the main differential geometric concepts and formulae remain valid. The extended coordinates and the extended frame, again denoted by $x^a$ and $c_{AB}$ satisfy the same defining equations and the extended fields, denoted again by $h$, $\phi^\alpha$ satisfy the field equations as before.

The \emph{null cone at} $q$ is defined as the set
\begin{equation*}
{\cal N}_q=\{p\in N_c|\Gamma(p)=0\}
\end{equation*}
where $\Gamma=\delta_{ab}x^ax^b$. It is the cone swept out by the complex null geodesics through $q$. For the type of systems we consider one very specific and important feature of this cone is that it is the characteristic cone not only for the Ricci operator in a harmonic gauge but also for the Laplace operators that act on the scalar fields. The coincidence of the characteristic cones for all the unknowns is due to the special form of the principal symbol of the system, it is diagonal and all the diagonal elements are the same. If this were not the case then the construction used in the present work would not be possible. It turns out for our problem that knowing the null data is equivalent to knowing the restriction of the holomorphic functions $\phi^\alpha$ to the null cone. The identification is made explicit later on.

Our problem can thus be formulated as the boundary value problem for the field equations with data given on the null cone by the functions $\phi^\alpha|_{{\cal N}_q}$. The difficulty with this formulation is that ${\cal N}_q$ is not a smooth hypersurface but an analytic set with a vertex at the point $q$, therefore an adapted coordinate and frame field is needed. The construction is done rigorously in \cite{Friedrich07} in terms of the principal bundle of spin frames over $N_c$, $SL(N_c)$. There a three-dimensional submanifold $\hat{N}$ of $SL(N_c)$ is constructed in such a way that together with the projection map $\pi:SL(N_c)\rightarrow N_c$ it induces coordinates on $N_c$. Here we describe how the construction is seen on $N_c$.

We start by taking the null vector $c_{11}$ at $q$ and constructing the null geodesic $\gamma$ with affine parameter $w$ whose tangent vector at $q$ is $c_{11}$. That is, $\gamma(w)$ is the affinely parametrized null geodesic that has $\gamma(0)=q$ and $\gamma'(0)=c_{11}$.

We define now a family of frames $e_{AB}$ at $q$ in terms of the frame $c_{AB}$ and a parameter $v\in\mathbb{C}$ in the following way
\begin{equation*}
e_{00}=c_{00}+2vc_{01}+v^2c_{11},\,\,\,e_{01}(v)=c_{01}+vc_{11},\,\,\,e_{11}(v)=c_{11}.
\end{equation*}
These frames are orthonormal in the sense that $h(e_{AB},e_{CD})=h_{ABCD}$. As $v$ varies $e_{00}$ covers all null directions at $q$ except $c_{11}$.

We parallelly propagate the frames $e_{AB}(v)$ over $\gamma(w)$. So now at each point of $\gamma(w)$ we have a family of frames $e_{AB}(v,w)$.

Consider a fixed value for $v=v_0$ and $w=w_0$. At $\gamma(w_0)$ we construct the null geodesic that goes through this point and has tangent vector $e_{00}(v_0,w_0)$ at $\gamma(w_0)$. We call $u$ the affine parameter on this null geodesic that vanishes at $\gamma(w_0)$ and we assign to a point on this geodesic the corresponding value of $u$ and the values $v=v_0$ and $w=w_0$. We also parallelly propagate the frame $e_{AB}(v_0,w_0)$ over the null geodesic. Doing this for all possible values of $v$ and $w$ covers all of $N_c$ and defines the frame field $e_{AB}$ there.

The functions $z^1=u$, $z^2=v$, $z^3=w$ define holomorphic coordinates on $\hat{N}$. We denote again $\pi$ the restriction of the projection to $\hat{N}$. The map $\pi$ induces a biholomorphic diffeomorphism of $\hat{N}'\equiv\hat{N}\backslash U_0$, where $U_0\equiv\{u=0\}$, onto $\pi(\hat{N}')$, but the gauge is singular in the sense that the set $U_0$, which is a two-surface on $\hat{N}$, projects into the curve $\gamma$ on $N_c$. We also need to distinguish the set $I\equiv\{u=0,w=0\}$, where $\pi(I)=q$. Finally, the set $W_0\equiv\{w=0\}$ projects onto ${\cal N}_q\backslash\gamma$ and will therefore define the initial data set for our problem.

It is important to recall that although the coordinates $z^a$ are well adapted to the geometry of the problem we are dealing with, and are holomorphic on $\hat{N}$, they are not a good coordinate system in all of $N_c$. The set $\{u=0\}$ corresponds to the curve $\gamma$ instead of being a hypersurface on $N_c$, and thus there is no way of assigning a $v$ coordinate to points on $\gamma$. For this the null curve $\gamma$ will be referred to as the \emph{the singular generator of ${\cal N}_q$ in the gauge determined by the frame $c_{AB}$ at $q$}. The singularity of the gauge is reflected in the following properties, deduced in \cite{Friedrich07}, and which are important for the present work.

Regarding the frame field, as seen on $\hat{N}$, if one writes $e_{AB}=e^a\,_{AB}\partial_{z^a}$, then on $\hat{N}'$,
\begin{equation*}
(e^a\,_{AB})=\left(\begin{array}{ccc} 1 & e^1\,_{01} & e^1\,_{11} \\ 0 & e^2\,_{01} & e^2\,_{11} \\ 0 & 0 & 1 \end{array}\right)=\left(\begin{array}{ccc} 1 & {\cal O}(u^2) & {\cal O}(u^2) \\ 0 & \frac{1}{2u}+{\cal O}(u) & {\cal O}(u) \\ 0 & 0 & 1 \end{array}\right)\mbox{ as }u\rightarrow 0.
\end{equation*}
We shall write
\begin{equation*}
e^a\,_{AB}=e^{*a}\,_{AB}+\hat{e}^a\,_{AB},
\end{equation*}
with singular part
\begin{equation*}
e^{*a}\,_{AB}=\delta^a_1\epsilon_A\,^0\epsilon_B\,^0+\delta^a_2\frac{1}{u}\epsilon_{(A}\,^0\epsilon_{B)}\,^1+\delta^a_3\epsilon_A\,^1\epsilon_B\,^1,
\end{equation*}
and holomorphic functions $\hat{e}^a\,_{AB}$ on $\hat{N}$ which satisfy
\begin{equation}\label{FrameGauge}
\hat{e}^a_{AB}={\cal O}(u)\mbox{ as }u\rightarrow 0.
\end{equation}
The connection coefficients $\Gamma_{ABCD}=\Gamma_{(AB)(CD)}$ satisfy
\begin{equation*}
\Gamma_{00AB}=0\mbox{ on }\hat{N},\,\,\,\Gamma_{11AB}=0\mbox{ on }U_0,
\end{equation*}
and also
\begin{equation*}
\Gamma_{ABCD}=\Gamma^*_{ABCD}+\hat{\Gamma}_{ABCD},
\end{equation*}
with singular part
\begin{equation*}
\Gamma^*_{ABCD}=-\frac{1}{u}\epsilon_{(A}\,^0\epsilon_{B)}\,^1\epsilon_C\,^0\epsilon_D\,^0,
\end{equation*}
and holomorphic functions $\hat{\Gamma}_{ABCD}$ on $\hat{N}$ which satisfy
\begin{equation}\label{GammaGauge}
\hat{\Gamma}_{ABCD}={\cal O}(u)\mbox{ as }u\rightarrow 0.
\end{equation}

We need a couple more of definitions and properties to be able to deal with spinor functions on $\hat{N}$.

In general, a holomorphic spinor field $\psi$ on $N_c$ is represented on $SL(N_c)$ by a holomorphic spinor-valued function $\psi_{A_1...A_j}$, given by the components of $\psi$ in corresponding spin frame. We shall use the notation $\psi_k=\psi_{(A_1...A_j)_k},k=0,..,j$, where $(......)_k$ denotes the operation \emph{`symmetrize and set $k$ indices equal to $1$ the rest equal to $0$'}. These functions completely specify $\psi$ if $\psi$ is symmetric. They are then referred to as the \emph{essential components of $\psi$}.

As the induced map $\pi$ of $\hat{N}$ into $N_c$ is singular on $U_0$, not every holomorphic function of the $z^a$ can arise as a pull-back to $\hat{N}$ of a holomorphic function on $N_c$. The former must have a special type of expansion in terms of the $z^a$ which reflects the particular relation between the `angular' coordinate $v$ and the `radial' coordinate $u$. The following definition and lemma, taken from \cite{Friedrich07}, are needed for manipulating spinor functions on $\hat{N}$.
\begin{defin}
A holomorphic function $g$ on $\hat{N}$ is said to be of $v$-finite expansion type $k_g$, with $k_g$ an integer, if it has in terms of the coordinates $u$, $v$, and $w$ a Taylor expansion at the origin of the form
\begin{equation*}
g=\sum_{p=0}^\infty\sum_{m=0}^\infty\sum_{n=0}^{2m+k_g}g_{m,n,p}u^mv^nw^p
\end{equation*}
where it is assumed that $g_{m,n,p}=0$ if $2m+k_g<0$.
\end{defin}
\begin{lem}\label{expansionType}
Let $\phi_{A_1...A_j}$ be a holomorphic, symmetric, spinor-valued function on $SL(N_c)$. Then the restrictions of its essential components $\phi_k=\phi_{(A_1...A_j)_k}$, $0\leq k\leq j$, to $\hat{N}$ satisfy
\begin{equation*}
\partial_v\phi_k=(j-k)\phi_{k+1},\,\,\,k=0,...,j,\mbox{ on }U_0,
\end{equation*}
(where we set $\phi_{j+1}=0$) and $\phi_k$ is of expansion type $j-k$.
\end{lem}

As stated at the beginning of this section, prescribing the null data is equivalent to knowing $\phi^\alpha|_{{\cal N}_q}$. Following \cite{Friedrich07} it is possible to see how this fit into our particular gauge. Consider the normal frame $c_{AB}$ on $N_c$ near $q$ and denote the null data of $h$ in this frame by
\begin{equation*}
{\cal D}^{\alpha*}=\{D_{(A_pB_p}...D_{A_1B_1)}\phi^\alpha(q),\hspace{0.5cm}p=1,2,3,...\},
\end{equation*}
then on $W_0$
\begin{equation}\label{expansionPhi}
 \phi^\alpha(u,v)=\sum_{m=0}^\infty\sum_{n=0}^{2m}\psi^\alpha_{m,n}u^mv^n,
\end{equation}
where
\begin{equation*}
\psi^\alpha_{m,n}=\frac{1}{m!}
\left(\begin{array}{c} 2m \\ n \end{array}\right)D_{(A_mB_m}...D_{A_1B_1)_n}\phi^\alpha(q),\,\,\,0\leq n\leq 2m.
\end{equation*}
This shows how to determine $\phi^\alpha(u,v)$ from the null data ${\cal D}^{\alpha*}$ and vice versa.

\section{The field equations on $\hat{N}$}\label{sectionConformalEquations}

Now, having the coordinates and frame field, we can use the frame calculus in its standard form. Given the fields $\phi^\alpha$ and using the frame $e_{AB}$ and the connection coefficients $\Gamma_{ABCD}$ on $\hat{N}$, we set
\begin{eqnarray*}
r_{ABCDEF} & \equiv & e_{CD}(\Gamma_{EFAB})-e_{EF}(\Gamma_{CDAB})+\Gamma_{EF}\,^{K}\,_{C}\Gamma_{DKAB}\\
&& +\Gamma_{EF}\,^{K}\,_{D}\Gamma_{CKAB}-\Gamma_{CD}\,^{K}\,_{E}\Gamma_{KFAB}-\Gamma_{CD}\,^{K}\,_{F}\Gamma_{EKAB}\\
&& +\Gamma_{EF}\,^{K}\,_{B}\Gamma_{CDAK}-\Gamma_{CD}\,^{K}\,_{B}\Gamma_{EFAK}-t_{CD}\,^{GH}\,_{EF}\Gamma_{GHAB}
\end{eqnarray*}
and define there the quantities $t_{AB}\,^{EF}\,_{CD}$, $R_{ABCDEF}$, $\sigma^\alpha_{AB}$, $\Sigma^\alpha_{AB}$ by
\begin{equation*}
\hspace{-2cm}t_{AB}\,^{EF}\,_{CD}e^{a}\,_{EF}\equiv2\Gamma_{AB}\,^{E}\,_{(C}e^{a}\,_{D)E}-2\Gamma_{CD}\,^{E}\,_{(A}e^{a}\,_{B)E}-e^{a}\,_{CD,b}e^{b}\,_{AB}+e^{a}\,_{AB,b}e^{b}\,_{CD},
\end{equation*}
\begin{eqnarray*}
R_{ABCDEF} & \equiv & r_{ABCDEF}-\frac{1}{2}\left[\left(S_{ABCE}-\frac{1}{6}R h_{ABCE}\right)\epsilon_{DF}\right.\\
&& \left.+\left(S_{ABDF}-\frac{1}{6}R h_{ABDF}\right)\epsilon_{CE}\right],
\end{eqnarray*}
\begin{equation*}
\sigma^\alpha_{AB}\equiv D_{AB}\phi^\alpha-\phi^\alpha_{AB},
\end{equation*}
\begin{equation*}
\Sigma^\alpha_{AB}\equiv D^{P}\,_{A}\phi_{BP}+\frac{1}{2}\epsilon_{AB}f^\alpha.
\end{equation*}
It is important to note that $R$ and $S_{ABCD}$, defined in \ref{spinorR}, \ref{spinorS}, represent functions of the fields that need not bear relation with the geometric Ricci scalar and trace-free part of the Ricci tensor. The equality among the quantities defined in terms of the fields $\phi^\alpha$ and its derivatives and the geometric quantities comes when the field equations are solved.

The tensor fields on the left hand side have been introduced as labels for the equations and for discussing in an ordered manner their interdependencies. In terms of these tensor fields, the field equations read
\begin{equation*}
 t_{AB}\,^{EF}\,_{CD}e^{a}\,_{EF}=0,\hspace{0.5cm}R_{ABCDEF}=0,\hspace{0.5cm}\sigma^\alpha_{AB}=0,\hspace{0.5cm}\Sigma^\alpha_{AB}=0.
\end{equation*}
The first equation is Cartan's first structural equation with the requirement that the metric connection be torsion free. The second equation is equation \ref{ricciSpinor}. The third equation define the symmetric spinors $\phi^\alpha_{AB}$. The last equation is the field equation \ref{lapPhiSpinor} in therms of $\phi^\alpha_{AB}$.

We want to see how to calculate in our particular gauge a formal expansion of the fields using the initial data in the form $\phi^\alpha(u,v)$. As the system of field equations is overdetermined we have to choose a subsystem of it. In the rest of this section we choose a particular subsystem, writing the chosen equations in our gauge, and at the end we see how a formal expansion is determined by these equations and the initial data.

\subsection{The $\sigma^\alpha_{00}=0$ equations}
The first set of equations that needs particular attention are the equations $\sigma^\alpha_{00}=0$. In our gauge they read
\begin{equation*}
\partial_{u}\phi^\alpha=\phi^\alpha_{00}.
\end{equation*}
These equations are used in the following to calculate $\phi^\alpha_{00}$ each time we know $\phi^\alpha$ as a function of $u$. In particular, as $\phi^\alpha$ will be prescribed on $W_0$ as part of the initial data, this equation allows us to calculate $\phi^\alpha_{00}$ there immediately.

\subsection{The `$\partial_u$-equations'}\label{dUequations}
We now present what we will refer to as the `$\partial_u$-equations'. These equations are chosen because they have the following features. They are a system of PDE's for the set of functions $\hat{e}^a\,_{A1}$, $\hat{\Gamma}_{A1CD}$ and $\phi^\alpha_{A1}$, which comprise all the unknowns with the exceptions of the free data $\phi^\alpha$ and the derived functions $\phi^\alpha_{00}$. They are all interior equations on the hypersurfaces $\{w=w_0\}$ in the sense that only derivatives in the directions of $u$ and $v$ are involved, in particular, if we consider the hypersurface $W_0$, they are all inner equations in ${\cal N}_q$. The possibility of choosing such a subsystem of inner equations in ${\cal N}_q$ is due to the coincidence of the characteristic cones for all the unknowns. Also they split into a hierarchy that will be presented in the following section. The `$\partial_u$-equations' are:

\noindent Equations $t_{AB}\,^{EF}\,_{00}e^{a}\,_{EF}=0:$
\begin{eqnarray*}
&& \partial_{u}\hat{e}^{1}\,_{01}+\frac{1}{u}\hat{e}^{1}\,_{01}=-2\hat{\Gamma}_{0101}+2\hat{\Gamma}_{0100}\hat{e}^{1}\,_{01},\\
&& \partial_{u}\hat{e}^{2}\,_{01}+\frac{1}{u}\hat{e}^{2}\,_{01}=\frac{1}{u}\hat{\Gamma}_{0100}+2\hat{\Gamma}_{0100}\hat{e}^{2}\,_{01},\\
&& \partial_{u}\hat{e}^{1}\,_{11}=-2\hat{\Gamma}_{1101}+2\hat{\Gamma}_{1100}\hat{e}^{1}\,_{01},\\
&& \partial_{u}\hat{e}^{2}\,_{11}=\frac{1}{u}\hat{\Gamma}_{1100}+2\hat{\Gamma}_{1100}\hat{e}^{2}\,_{01}.
\end{eqnarray*}
Equations $R_{AB00EF}=0:$
\begin{eqnarray*}
&& \partial_{u}\hat{\Gamma}_{0100}+\frac{2}{u}\hat{\Gamma}_{0100}-2\hat{\Gamma}^{2}_{0100}=\frac{1}{2}\big(F^\alpha\partial_u\phi^\alpha_{00}+F^{\alpha\beta}\phi^\alpha_{00}\phi^\beta_{00}\big),\\
&& \partial_{u}\hat{\Gamma}_{0101}+\frac{1}{u}\hat{\Gamma}_{0101}-2\hat{\Gamma}_{0100}\hat{\Gamma}_{0101}=\frac{1}{2}\big(F^\alpha\partial_u\phi^\alpha_{01}+F^{\alpha\beta}\phi^\alpha_{00}\phi^\beta_{01}\big),\\
&& \partial_{u}\hat{\Gamma}_{0111}+\frac{1}{u}\hat{\Gamma}_{0111}-2\hat{\Gamma}_{0100}\hat{\Gamma}_{0111}=\frac{1}{2}\big(F^\alpha\partial_u\phi^\alpha_{11}+F^{\alpha\beta}\phi^\alpha_{00}\phi^\beta_{11}+\tilde{f}\big)-\frac{1}{12}\hat{f},\\
&& \partial_{u}\hat{\Gamma}_{1100}+\frac{1}{u}\hat{\Gamma}_{1100}-2\hat{\Gamma}_{0100}\hat{\Gamma}_{1100}=F^\alpha\partial_u\phi^\alpha_{01}+F^{\alpha\beta}\phi^\alpha_{00}\phi^\beta_{01},\\
&& \partial_{u}\hat{\Gamma}_{1101}-2\hat{\Gamma}_{1100}\hat{\Gamma}_{0101}=F^\alpha\partial_u\phi^\alpha_{11}+F^{\alpha\beta}\phi^\alpha_{00}\phi^\beta_{11}+\tilde{f}+\frac{1}{12}\hat{f},\\
&& \partial_{u}\hat{\Gamma}_{1111}-2\hat{\Gamma}_{1100}\hat{\Gamma}_{0111}=F^\alpha\Big(\frac{1}{2u}\partial_{v}\phi^\alpha_{11}+\hat{e}^{1}\,_{01}\partial_{u}\phi^\alpha_{11}+\hat{e}^{2}\,_{01}\partial_{v}\phi^\alpha_{11}\\
&&-2\hat{\Gamma}_{0111}\phi^\alpha_{01}+2\hat{\Gamma}_{0101}\phi^\alpha_{11}\Big)+F^{\alpha\beta}\phi^\alpha_{01}\phi^\beta_{11}.
\end{eqnarray*}
Equations $\Sigma^\alpha_{A0}=0:$
\begin{equation*}
\hspace{-1.5cm}\partial_{u}\phi^\alpha_{01}=\frac{1}{2u}(\partial_{v}\phi^\alpha_{00}-2\phi^\alpha_{01})+\hat{e}^{1}\,_{01}\partial_{u}\phi^\alpha_{00}+\hat{e}^{2}\,_{01}\partial_{v}\phi^\alpha_{00}-2\hat{\Gamma}_{0101}\phi^\alpha_{00}+2\hat{\Gamma}_{0100}\phi^\alpha_{01},
\end{equation*}
\begin{equation*}
\hspace{-1.5cm}\partial_{u}\phi^\alpha_{11}=\frac{1}{2u}\left(\partial_{v}\phi^\alpha_{01}-\phi^\alpha_{11}\right)+\hat{e}^{1}\,_{01}\partial_{u}\phi^\alpha_{01}+\hat{e}^{2}\,_{01}\partial_{v}\phi^\alpha_{01}-\hat{\Gamma}_{0111}\phi^\alpha_{00}+\hat{\Gamma}_{0100}\phi^\alpha_{11}+\frac{1}{2}f^\alpha.
\end{equation*}

\subsection{The $\partial_u$-equations hierarchy}\label{hierarchy}

The system of $\partial_u$-equations splits into a hierarchy of subsystems as follows\\
H1) $R_{000001}=0$,\\
H2) $t_{01}\,^{EF}\,_{00}e^2\,_{EF}=0$,\\
H3) $t_{01}\,^{EF}\,_{00}e^1\,_{EF}=0$, $R_{010001}=0$, $\Sigma^\alpha_{00}=0$,\\
H4) $R_{000011}=0$,\\
H5) $t_{11}\,^{EF}\,_{00}e^2\,_{EF}=0$,\\
H6) $R_{110001}=0$, $\Sigma^\alpha_{10}=0$,\\
H7) $R_{010011}=0$,\\
H8) $R_{110011}=0$,\\
H9) $t_{11}\,^{EF}\,_{00}e^1\,_{EF}=0$.

The defining property of the hierarchy is the following. If $\phi^\alpha$ are prescribed on $\{w=w_0\}$ and using $\sigma^\alpha_{00}=0$, then H1 reduces to an ODE. Once we have its solution, H2 reduces to an ODE. Given its solution, H3 reduces to a system of ODE's, with coefficients that are calculated by operations interior to $\{w=w_0\}$ from the previously known or calculated functions. This procedure continues till the end of the hierarchy. So, given $\phi^\alpha$ on $\{w=w_0\}$ and the appropriate initial data on $U_0\cap\{w=w_0\}$, all the unknowns can be determined on $\{w=w_0\}$ by solving a sequence of ODE's in the independent variable $u$.

\subsection{The `$\partial_w$-equations'}
The initial data, $\phi^\alpha$ are prescribed on $W_0$, and to determine their evolution off $W_0$ we need the equations $\sigma^\alpha_{11}=0$, referred to as the $\partial_w$-equations, which read
\begin{equation*}
 \partial_w\phi^\alpha=\phi^\alpha_{11}-\hat{e}^1\,_{11}\partial_u\phi^\alpha-\hat{e}^2\,_{11}\partial_v\phi^\alpha.
\end{equation*}

\subsection{Calculating the formal expansion}

Let us call $X$ any of the unknowns that we are solving for, i.e. $\hat{e}^a_{AB}$, $\hat{\Gamma}_{ABCD}$, $\phi^\alpha$, $\phi^\alpha_{AB}$. We want to see that it is possible to obtain a formal expansion of $X$ in a neighbourhood of $q$ using the $\sigma^\alpha_{00}$-equations, the $\partial_u$-equations and the $\partial_w$-equations, given $\phi^\alpha$ on $W_0$ as our datum. We give here an inductive argument showing that with our setting $\partial_w^kX|_{W_0}$ can be determined for all $k$.

We need the following conditions, obtained from the gauge requirements \ref{FrameGauge}, \ref{GammaGauge}
\begin{eqnarray*}
&& \partial^k_{w}\hat{e}^a\,_{A1}|_{I}=0,\hspace{1cm}a=1,2,\hspace{1cm}A=0,1,\hspace{1cm}k\geq 0,\\
&& \partial^k_{w}\hat{\Gamma}_{A1CD}|_{I}=0,\hspace{1cm}A,C,D=0,1,\hspace{1cm}k\geq 0,
\end{eqnarray*}
and from the $\sigma^\alpha_{00}=0$ equations and the spinorial behaviour as discussed in Lemma \ref{expansionType},
\begin{equation*}
\partial^k_{w}\phi_{A1}|_{I}=\frac{1}{2}\partial_{u}\partial^{1+A}_{v}\partial^k_{w}\phi|_{I},\hspace{1cm}A=0,1,\hspace{1cm}k\geq0,
\end{equation*}
which are our initial conditions for the $\partial_w^k$-derivatives of the $\partial_u$-equations.

Using the initial conditions for $k=0$ and following what has been said in Subsection \ref{hierarchy} we successively integrate the subsystems H1 to H9 to determine all of $X$ on $W_0$.

As inductive hypothesis we assume as known $\partial^p_{w}X|_{W_{0}},\,\,\,0\leq p \leq k-1,\,\,\,k\geq 1$. Applying formally $\partial_w^{k-1}$ to the $\partial_w$-equations, and restricting them to $W_0$, we find $\partial^k_{w}\phi^\alpha|_{W_{0}}$ in terms of known functions. We apply formally $\partial_w^{k}$ to the $\partial_u$-equations. This is a system of PDE's where the unknowns are $\partial_w^kX$. Keeping the discussed hierarchy and considering the functions that we already know on $W_0$, it again becomes a sequence of ODE's, which together with the initial conditions on $I$ can be integrated on $W_0$. Thus we know $\partial^k_{w}X|_{W_{0}}$ and the induction step is completed.

The procedure just stated shows that we know $\partial^k_{w}X|_{W_{0}}$ for all $k$. Expanding these functions around $q=\{u=0,v=0,w=0\}$ gives
\begin{equation*}
\partial_u^m\partial_v^n\partial_w^pX|_q\hspace{1cm}\forall\,\,m,n,p,
\end{equation*}
and the procedure gives a unique sequence of expansion coefficients for all the functions in $X$.
\begin{lem}\label{expansionCoefficients}
The procedure described above determines at the point $O=(u=0,v=0,w=0)$ from the data $\phi^\alpha$ given on $W_0$ according to \ref{expansionPhi} a unique sequence of expansion coefficients
\begin{equation*}
\partial_u^m\partial_v^n\partial_w^pX(O),\hspace{1cm}m,n,p=0,1,2,...
\end{equation*}
where $X$ stands for any of the functions $\hat{e}^a\,_{AB}$, $\hat{\Gamma}_{ABCD}$, $\phi^\alpha$, $\phi^\alpha_{AB}$.

If the corresponding Taylor series are absolutely convergent in some neighbourhood $Q$ of $O$, they define a solution to the equation $\sigma^\alpha_{00}=0$, to the $\partial_u$-equations and to the $\partial_w$-equations on $Q$.
\end{lem}

By Lemma \ref{expansionType} we know that all spinor-valued functions should have a specific $v$-finite expansion type. The following lemma, whose proof is quite similar to the proof in \cite{Friedrich07}, is important for handling the estimates in the following section.
\begin{lem}\label{expansionTypeFields}
If the data $\phi^\alpha$ are given on $W_0$ as in \ref{expansionPhi} the formal expansions of the fields obtained in Lemma \ref{expansionCoefficients} correspond to ones of functions of $v$-finite expansion types given by
\begin{equation*}
 \begin{array}{l l l}
 k_{\hat{e}^1\,_{AB}}=-A-B, & k_{\hat{e}^2\,_{AB}}=3-A-B, & AB=01,11,\\
 k_{\hat{\Gamma}_{01AB}}=2-A-B, & k_{\hat{\Gamma}_{11AB}}=1-A-B, & A,B=0,1,\\
 k_\phi^\alpha=0, & k_{\phi^\alpha_{AB}}=2-A-B, & A,B=0,1.
 \end{array}
\end{equation*}
\end{lem}

\section{Convergence of the formal expansion}\label{convergence}
In the previous section we have seen how to calculate a formal expansion for $\hat{e}^a\,_{AB}$, $\hat{\Gamma}_{ABCD}$, $\phi^\alpha$, $\phi^\alpha_{AB}$ given $\phi^\alpha|_{W_0}$, or, what is the same, given the null data. From Lemma \ref{estimatesNullData} we know which are the necessary conditions on the null data in order to have analytic solutions of the conformal field equations. In this section we show that those conditions are also sufficient for the formal expansion determined in the previous section to be absolutely convergent.

We consider the abstract null data as given by $n$ sequences
\begin{equation*}
\hat{\cal D}^\alpha=\{\psi^\alpha_{A_1B_1},\psi^\alpha_{A_2B_2A_1B_1},\psi^\alpha_{A_3B_3A_2B_2A_1B_1},...\}
\end{equation*}
of totally symmetric spinors satisfying the reality condition \ref{realityCond} and we construct $\phi^\alpha|_{W_0}$ by setting in the expansions \ref{expansionPhi}
\begin{equation*}
D_{(A_mB_m}...D_{A_1B_1)}\phi^\alpha(q)=\psi^\alpha_{A_mB_m...A_1B_1},\hspace{1cm}m\geq 0.
\end{equation*}
Observing Lemma \ref{estimatesNullData} one finds as a necessary condition for the functions $\phi^\alpha$ on $W_0$ to determine an analytic solution to the conformal static vacuum field equations that its non-vanishing Taylor coefficients at the point $O$ satisfy estimates of the form
\begin{equation}\label{firstEstimatePhi}
|\partial_u^m\partial_v^n\phi^\alpha(O)|\leq
\left(\begin{array}{c} 2m \\ n \end{array}\right)m!n!\frac{M}{r^m},\hspace{1cm}m\geq 0,\hspace{1cm}0\leq n\leq 2m.
\end{equation}
This conditions are also sufficient for $\phi^\alpha(u,v)$ to be holomorphic functions on $W_0$. So the null data give rise to $n$ analytic functions $\phi^\alpha$ on $W_0$.

From $\sigma^\alpha_{00}=0$ we have $\phi^\alpha_{00}=\partial_u\phi^\alpha$, so having $\phi^\alpha|_{W_0}$ we have $\phi^\alpha_{00}|_{W_0}$, which is also an analytic function on $W_0$.

Following Lemma 6.1 in \cite{Friedrich07}, we can derive from \ref{firstEstimatePhi} slightly different type of estimates for $\phi^\alpha(u,v)$ which are more convenient in our case.
\begin{lem}
Let $e$ be the Euler number. For given $\rho_{\phi^\alpha}$ in $\mathbb{R}$, such that $0<\rho_{\phi^\alpha}<e^2$, there exist positive constants $c_{\phi^\alpha}$, $r_{\phi^\alpha}$ so that \ref{firstEstimatePhi} imply estimates of the form
\begin{equation}\label{EstimatePhi}
|\partial^m_u\partial^n_v\phi^\alpha(O)|\leq c_{\phi^\alpha} \frac{r_{\phi^\alpha}^m m! \rho^n_{\phi^\alpha} n!}{(m+1)^2 (n+1)^2},\hspace{1cm}m\geq 0,\hspace{1cm}0\leq n\leq 2m.
\end{equation}
\end{lem}
We present our estimates.
\begin{lem}\label{mainEstimates}
Assume $\phi^\alpha=\phi^\alpha(u,v)$ are holomorphic functions defined on some open neighbourhood $U$ of $O=\{u=0,v=0,w=0\}$ in $W_0=\{w=0\}$ which have expansions of the form
\begin{equation*}
\phi^\alpha(u,v)=\sum_{m=0}^\infty\sum_{n=0}^{2m}\psi^\alpha_{m,n}u^mv^n
\end{equation*}
so that its Taylor coefficients at the point $O$ satisfy estimates of the type \ref{EstimatePhi} with some positive constants $c_{\phi^\alpha}$, $r_{\phi^\alpha}$, and $\rho_{\phi^\alpha}<\frac{1}{8}$. Then there exist positive constants $r$, $\rho$, $c_{\hat{e}^a\,_{AB}}$, $c_{\hat{\Gamma}_{ABCD}}$, $c_{\phi^\alpha_{AB}}$ so that the expansion coefficients determined from $\phi^\alpha$ in Lemma \ref{expansionCoefficients} satisfy for $m,n,p=0,1,2,...$
\begin{equation}\label{convergenceEstimates}
|\partial_u^m\partial_v^n\partial_w^pX(O)|\leq c_X \frac{r^{m+p+q_X}(m+p)!\rho^nn!}{(m+1)^2(n+1)^2(p+1)^2}
\end{equation}
where $X$ stands for any of the functions $\hat{e}^a\,_{AB}$, $\hat{\Gamma}_{ABCD}$, $\phi^\alpha$, $\phi^\alpha_{AB}$ and
\begin{equation*}
q_{\hat{e}^a\,_{AB}}=q_{\hat{\Gamma}_{ABCD}}=-1,\hspace{1cm}q_{\phi^\alpha}=q_{\phi^\alpha_{AB}}=0.
\end{equation*}
\end{lem}
\begin{remark}
Taking into account the $v$-finite expansion types of the functions $X$ obtained in Lemma \ref{expansionTypeFields}, we can replace the right hand sides in the estimates above by zero if $n$ is large enough relative to $m$. This will not be pointed out at each step and for convenience the estimates will be written as above.
\end{remark}

We take the following four lemmas from \cite{Friedrich07}. The first states the necessary part of the estimates, and the other three are needed in order to manipulate the estimates in the proof of Lemma \ref{mainEstimates}.
\begin{lem}
If $g$ is a holomorphic function near $O$, then there exist positive constants $c$, $r_0$, $\rho_0$ such that
\begin{equation*}
|\partial_u^m\partial_v^n\partial_w^pg(O)|\leq c \frac{r^{m+p}(m+p)!\rho^nn!}{(m+1)^2(n+1)^2(p+1)^2},\,\,\,m,n,p=0,1,2,...
\end{equation*}
for any $r\geq r_0$, $\rho\geq\rho_0$. If in addition $g(0,v,0)=0$, the constants can be chosen such that
\begin{equation*}
|\partial_u^m\partial_v^n\partial_w^pg(O)|\leq c \frac{r^{m+p-1}(m+p)!\rho^nn!}{(m+1)^2(n+1)^2(p+1)^2},\,\,\,m,n,p=0,1,2,...
\end{equation*}
for any $r\geq r_0$, $\rho\geq\rho_0$.
\end{lem}
\begin{lem}
For any non-negative integer $n$ there is a positive constant $C$, $C>1$, independent of $n$ so that
\begin{equation*}
\sum_{k=0}^n\frac{1}{(k+1)^2(n-k+1)^2}\leq C\frac{1}{(n+1)^2}.
\end{equation*}
\end{lem}
In the following $C$ will always denote the constant above.
\begin{lem}
For any integers $m$, $n$, $k$, $j$, with $0\leq k\leq m$, and $0\leq j\leq n$ resp. $0\leq j \leq n-1$ holds
\begin{equation*}
\left(\begin{array}{c} m \\ k \end{array}\right)\left(\begin{array}{c} n \\ j \end{array}\right)\leq \left(\begin{array}{c} m+n \\ k+j \end{array}\right)\mbox{   resp.   }\left(\begin{array}{c} m \\ k \end{array}\right)\left(\begin{array}{c} n-1 \\ j \end{array}\right)\leq \left(\begin{array}{c} m+n \\ k+j \end{array}\right).
\end{equation*}
\end{lem}
\begin{lem}\label{multiplEstimates}
Let $m$, $n$, $p$ be non-negative integers and $g_i$, $i=1,...,N$, be smooth complex valued functions of $u$, $v$, $w$ on some neighbourhood $U$ of $O$ whose derivatives satisfy on $U$ (resp. at a given point $q\in U$) estimates of the form
\begin{equation*}
|\partial_u^j\partial_v^k\partial_w^lg_i(O)|\leq c_i \frac{r^{j+l+q_i}(j+l)!\rho^kk!}{(j+1)^2(k+1)^2(l+1)^2}
\end{equation*}
for $0\leq j\leq m$, $0\leq k\leq n$, $0\leq l\leq p$, with some positive constants $c_i$, $r$, $\rho$ and some fixed integers $q_i$ (independent of $j$, $k$, $l$). Then one has on $U$ (resp. at $q$) the estimates
\begin{equation}\label{estimateMultiplication}
\hspace{-1.5cm}|\partial_u^m\partial_v^n\partial_w^p(g_1\cdot ... \cdot g_N)(O)|\leq C^{3(N-1)}c_1\cdot ... \cdot c_N \frac{r^{m+p+q_1+...+q_N}(m+p)!\rho^nn!}{(m+1)^2(n+1)^2(p+1)^2}.
\end{equation}
\end{lem}
\begin{remark}
This lemma remains true if $m$, $n$, $p$ are replaced in \ref{estimateMultiplication} by integers $m'$, $n'$, $p'$ with $0\leq m'\leq m$, $0\leq n'\leq n$, $0\leq p'\leq p$.

The factor $C^{3(N-1)}$ in \ref{estimateMultiplication} can be replaced by $C^{(3-s)(N-1)}$ if $s$ of the integers $m$, $n$, $p$ vanish.
\end{remark}
From now on we consider that a function in a modulus sign is evaluated at the origin $O$.

\begin{proof}[Proof of Lemma \ref{mainEstimates}]
The proof is by induction, following the inductive procedure which led to Lemma \ref{expansionCoefficients}. A general outline is as follows. We start leaving the choice of the constants $r$, $\rho$, $c_f$ open. We use the induction hypothesis and the equations that lead to Lemma \ref{expansionCoefficients} to derive estimates for the derivatives of the next order. These estimates are of the form
\begin{equation}\label{generalEstimateForm}
|\partial_u^m\partial_v^n\partial_w^pX|\leq c_X \frac{r^{m+p+q_X}(m+p)!\rho^nn!}{(m+1)^2(n+1)^2(p+1)^2}A_X
\end{equation}
with certain constants $A_X$ which depend on $m$, $n$, $p$ and the constants $c_X$, $r$ and $\rho$. Sometimes superscripts will indicate to which order of differentiability particular constants $A_X$ refer. In the way we will have to make assumptions on $r$ to proceed with the induction step. We shall collect these conditions and the constants $A_X$, or estimates for them, and at the end it will be shown that the constants $c_X$, $r$ and $\rho$ can be adjusted so that all conditions are satisfied and $A_X\leq 1$. This will complete the induction proof.

In order not to write long formulae that do not add to the understanding of the procedure, we state here some properties that are used to simplify the estimates:
\begin{itemize}
\item{During the procedure we need estimates on the derivatives of the functions $f$, $f^\alpha$, $F^\alpha$ and $F^{\alpha\beta}$, which are analytic functions of the fields $\phi^\alpha$ and $\phi^\alpha_{AB}$. The functions are also scalars if we consider them as functions on the manifold through their dependence on the fields. This and the Cauchy estimates for derivatives of analytic functions allow us to get the following result.

Let us denote by $g$ any of the functions $f$, $f^\alpha$, $F^\alpha$ and $F^{\alpha\beta}$. If, for $r\geq r_0$, $0\leq j\leq m$, $0\leq k\leq n$, $0\leq l\leq p$,
\begin{eqnarray*}
 |\partial_u^j\partial_v^k\partial_w^l\phi^\alpha|\leq c_{\phi^\alpha}\frac{r^{j+l}(j+l)!\rho^kk!}{(j+1)^2(k+1)^2(l+1)^2},\\
 |\partial_u^j\partial_v^k\partial_w^l\phi^\alpha_{AB}|\leq c_{\phi^\alpha_{AB}}\frac{r^{j+l}(j+l)!\rho^kk!}{(j+1)^2(k+1)^2(l+1)^2}
\end{eqnarray*}
then there exists positive constants $c_g$, $R_0\geq r_0$ such that for $R\geq R_0$
\begin{equation}\label{estimateg}
 |\partial_u^m\partial_v^n\partial_w^pg|\leq c_g\frac{R^{m+p}(m+p)!\rho^nn!}{(m+1)^2(n+1)^2(p+1)^2}.
\end{equation}
Using this for calculating the estimates leads to terms with the factor $\frac{R}{r}$. We can then fix $R$ by making $R=R_0$ and add to the requirements for $r$ that $r\geq R_0$. We have then
\begin{equation}\label{condR0}
 \frac{R}{r}\leq 1.
\end{equation}
}
\item{After calculating the estimates and using \ref{estimateg} and \ref{condR0} we find that all the $A$'s satisfy inequalities of the form
\begin{equation*}
A\leq\alpha+\frac{\hat{\alpha}}{r},
\end{equation*}
where $\alpha,\hat{\alpha}$ are constants that do not depend on $r$. If $\hat{\alpha}=0$ then we have to show that we can make $\alpha\leq 1$. If the $\hat{\alpha}$'s are not zero we can take a constant $a$, $0<a<1$, and require that $\alpha\leq a$ and then choose $r$ large enough such that $\frac{\hat{\alpha}}{r}\leq 1-a$. In the estimates that follows, we shall not write the explicit expressions for the $\hat{\alpha}$'s, as they do not play any role if we are able to make $r$ big enough at the end of the procedure.
}
\end{itemize}
As the $\phi^\alpha$'s are analytic functions of $u$ and $v$ on $W_0$, also the $\phi^\alpha_{00}$'s are. Then there exist positive constants $0<\rho_{\phi^\alpha},\rho_{\phi^\alpha_{00}}\leq \frac{1}{8}$ and $c_{\phi^\alpha},c_{\phi^\alpha_{00}},r_{\phi^\alpha},r_{\phi^\alpha_{00}}$ such that
\begin{eqnarray*}
 |\partial_u^m\partial_v^n\phi^\alpha| & \leq & c_{\phi^\alpha}\frac{r_{\phi^\alpha}^mm!\rho_{\phi^\alpha}^nn!}{(m+1)^2(n+1)^2},\\
 |\partial_u^m\partial_v^n\phi^\alpha_{00}| & \leq & c_{\phi^\alpha_{00}}\frac{r_{\phi^\alpha_{00}}^mm!\rho_{\phi^\alpha_{00}}^nn!}{(m+1)^2(n+1)^2}.
\end{eqnarray*}
As the inequalities do not change if the constants are changed for bigger constants, we choose (but leaving the precise value open)
\begin{equation*}
 r \geq \max\{r_{\phi^\alpha},r_{\phi^\alpha_{00}}\},\hspace{1cm}\rho \geq \max\{\rho_{\phi^\alpha},\rho_{\phi^\alpha_{00}}\},
\end{equation*}
then
\begin{eqnarray*}
 |\partial_u^m\partial_v^n\partial_w^0\phi^\alpha| & \leq & c_{\phi^\alpha}\frac{r^mm!\rho^nn!}{(m+1)^2(n+1)^2},\\
 |\partial_u^m\partial_v^n\partial_w^0\phi^\alpha_{00}| & \leq & c_{\phi^\alpha_{00}}\frac{r^mm!\rho^nn!}{(m+1)^2(n+1)^2}.
\end{eqnarray*}
Using how the frame fields and the coordinates where constructed,
\begin{eqnarray*}
 \hat{e}^a\,_{AB}|_{U_0}=0 & \Rightarrow & |\partial_u^0\partial_v^n\partial_w^p\hat{e}^a\,_{AB}|=0,\\
 \hat{\Gamma}_{ABCD}|_{U_0}=0 & \Rightarrow & |\partial_u^0\partial_v^n\partial_w^p\hat{\Gamma}_{ABCD}|=0.
\end{eqnarray*}
From the required spinorial behaviour we have
\begin{equation*}
 \phi_{01}=\frac{1}{2}\partial_v\phi_{00},\hspace{1cm}\phi_{11}=\partial_v\phi_{01},\hspace{1cm}\mbox{on }U_0,
\end{equation*}
then
\begin{equation*}
 |\partial_u^0\partial_v^n\partial_w^0\phi_{01}|=\frac{1}{2}|\partial_v^{n+1}\phi_{00}|\leq\frac{1}{2}c_{\phi^\alpha_{00}}\frac{\rho^{n+1}(n+1)!}{(n+2)^2}=c_{\phi^\alpha_{01}}\frac{\rho^nn!}{(n+1)^2}A_{\phi^\alpha_{01}}^{m=0,p=0},
\end{equation*}
where
\begin{equation*}
 A_{\phi^\alpha_{01}}^{m=0,p=0}=\frac{1}{2}\frac{c_{\phi^\alpha_{00}}}{c_{\phi^\alpha_{01}}}\rho\frac{(n+1)^3}{(n+2)^2}\leq\frac{1}{2}\frac{c_{\phi^\alpha_{00}}}{c_{\phi^\alpha_{01}}}\rho,
\end{equation*}
as $k_{\phi^\alpha_{01}}=1$ and then $n=0,1$.
Now that we have this inequality, and in the same way, we get
\begin{equation*}
 A_{\phi^\alpha_{11}}^{m=0,p=0}\leq\frac{1}{4}\frac{c_{\phi^\alpha_{01}}}{c_{\phi^\alpha_{11}}}\rho.
\end{equation*}
Now we have estimates for $\partial_u^0\partial_v^n\partial_w^0X$, with $n\in\mathbb{N}_0$. We assume as inductive hypothesis that we have estimates for $\partial_u^l\partial_v^n\partial_w^0X$, with $0\leq l\leq m-1$, $n\in\mathbb{N}_0$. We already have those estimates for $\phi^\alpha$ and $\phi^\alpha_{00}$. We use the $\partial_u$-equations to get estimates for $\partial_u^m\partial_v^n\partial_w^0X$ for the rest of the unknowns. The estimates obtained in this step are less restrictive than the estimates for general $p$, so we skip the enumeration of them and go on to the next step.

We assume that we have estimates for $\partial_u^m\partial_v^n\partial_w^lX$, with $0\leq l\leq p-1$, $m\in\mathbb{N}_0$, $n\in\mathbb{N}_0$, and use the system of equations and the properties stated on the lemmas of this section to get estimates for $\partial_u^m\partial_v^n\partial_w^pX$.

Using the $\partial_w$-equations we get
\begin{equation*}
 A_{\phi^\alpha}^{p\geq 1}\leq 4C^3c_{\hat{e}^1\,_{11}}+\frac{\hat{\alpha}_{\phi^\alpha}^{p\geq 1}}{r}.
\end{equation*}
Using the equation $D_{11}\phi^\alpha_{00}=D_{00}\phi^\alpha_{11}$, which follows from $\sigma^\alpha_{00}=0$ and $\sigma^\alpha_{11}=0$,
\begin{equation*}
A_{\phi^\alpha_{00}}^{p\geq 1}\leq \frac{4}{c_{\phi^\alpha_{00}}}\Big(c_{\phi^\alpha_{11}}+C^3c_{\hat{e}^1\,_{11}}c_{\phi^\alpha_{00}}\Big)+\frac{\hat{\alpha}_{\phi^\alpha_{00}}^{p\geq 1}}{r}.
\end{equation*}
For $m=0$ and $p\geq 1$ it is not possible to use the $\partial_u$-equations to get estimates, so it is necessary to use the spinorial behaviour to get, as before
\begin{equation*}
 A_{\phi^\alpha_{01}}^{m=0}\leq\frac{1}{2}\frac{c_{\phi^\alpha_{00}}}{c_{\phi^\alpha_{01}}}\rho,\hspace{1cm}A_{\phi^\alpha_{11}}^{m=0}\leq\frac{1}{4}\frac{c_{\phi^\alpha_{01}}}{c_{\phi^\alpha_{11}}}\rho.
\end{equation*}
Now we use the $\partial_u$-equations, obtaining
\begin{equation*}
 \begin{array}{l l}
  A_{\hat{e}^1\,_{01}}^{m\geq 1}\leq \frac{\hat{\alpha}_{\hat{e}^1\,_{01}}^{m\geq 1}}{r}, & A_{\hat{e}^2\,_{01}}^{m\geq 1}\leq \frac{c_{\hat{\Gamma}_{0100}}}{2c_{\hat{e}^2\,_{01}}}+\frac{\hat{\alpha}_{\hat{e}^2\,_{01}}^{m\geq 1}}{r},\\
  A_{\hat{e}^1\,_{11}}^{m\geq 1}\leq \frac{\hat{\alpha}_{\hat{e}^1\,_{11}}^{m\geq 1}}{r}, & A_{\hat{e}^2\,_{11}}^{m\geq 1}\leq \frac{c_{\hat{\Gamma}_{1100}}}{c_{\hat{e}^2\,_{11}}}+\frac{\hat{\alpha}_{\hat{e}^2\,_{11}}^{m\geq 1}}{r},\\
  A_{\hat{\Gamma}_{0100}}^{m\geq 1}\leq \frac{C^3c_{F^\alpha}c_{\phi^\alpha_{00}}}{2c_{\hat{\Gamma}_{0100}}}+\frac{\hat{\alpha}_{\hat{\Gamma}_{0100}}^{m\geq 1}}{r}, & A_{\hat{\Gamma}_{0101}}^{m\geq 1}\leq \frac{C^3c_{F^\alpha}c_{\phi^\alpha_{01}}}{2c_{\hat{\Gamma}_{0101}}}+\frac{\hat{\alpha}_{\hat{\Gamma}_{0101}}^{m\geq 1}}{r},\\
  A_{\hat{\Gamma}_{0111}}^{m\geq 1}\leq  \frac{C^3c_{F^\alpha}c_{\phi^\alpha_{11}}}{2c_{\hat{\Gamma}_{0111}}}+\frac{\hat{\alpha}_{\hat{\Gamma}_{0111}}^{m\geq 1}}{r}, & A_{\hat{\Gamma}_{1100}}^{m\geq 1}\leq \frac{C^3c_{F^\alpha}c_{\phi^\alpha_{01}}}{c_{\hat{\Gamma}_{1100}}}+\frac{\hat{\alpha}_{\hat{\Gamma}_{1100}}^{m\geq 1}}{r},\\
  A_{\hat{\Gamma}_{1101}}^{m\geq 1}\leq  \frac{C^3c_{F^\alpha}c_{\phi^\alpha_{11}}}{c_{\hat{\Gamma}_{1101}}}+\frac{\hat{\alpha}_{\hat{\Gamma}_{1101}}^{m\geq 1}}{r}, & A_{\hat{\Gamma}_{1111}}^{m\geq 1}\leq  \frac{C^3(\rho+C^3c_{\hat{e}^1\,_{01}})c_{F^\alpha}c_{\phi^\alpha_{11}}}{c_{\hat{\Gamma}_{1111}}}+\frac{\hat{\alpha}_{\hat{\Gamma}_{1111}}^{m\geq 1}}{r},\\
  A_{\phi^\alpha_{01}}^{m\geq 1}\leq \frac{(\rho+C^3c_{\hat{e}^1\,_{01}})c_{\phi^\alpha_{00}}}{c_{\phi^\alpha_{01}}}+\frac{\hat{\alpha}_{\phi^\alpha_{01}}^{m\geq 1}}{r}, & A_{\phi^\alpha_{11}}^{m\geq 1}\leq \frac{(\rho+C^3c_{\hat{e}^1\,_{01}})c_{\phi^\alpha_{01}}}{c_{\phi^\alpha_{11}}}+\frac{\hat{\alpha}_{\phi^\alpha_{11}}^{m\geq 1}}{r}.
 \end{array}
\end{equation*}

We take a constant $a$, $0<a<1$, whose precise value will be fixed later. We see that if we can make $r$ arbitrarily large, then all the $A$'s are less or equal to $1$ if the following inequalities are satisfied:
\begin{eqnarray*}
 \begin{array}{l l l}
  \frac{c_{\hat{\Gamma}_{0100}}}{2c_{\hat{e}^2\,_{01}}}\leq a, & \frac{c_{\hat{\Gamma}_{1100}}}{c_{\hat{e}^2\,_{11}}}\leq a, & \frac{C^3c_{F^\alpha}c_{\phi^\alpha_{00}}}{2c_{\hat{\Gamma}_{0100}}}\leq a,\\
  \frac{C^3c_{F^\alpha}c_{\phi^\alpha_{01}}}{2c_{\hat{\Gamma}_{0101}}}\leq a, & \frac{C^3c_{F^\alpha}c_{\phi^\alpha_{11}}}{2c_{\hat{\Gamma}_{0111}}}\leq a, & \frac{C^3c_{F^\alpha}c_{\phi^\alpha_{01}}}{c_{\hat{\Gamma}_{1100}}}\leq a,\\
  \frac{C^3c_{F^\alpha}c_{\phi^\alpha_{11}}}{c_{\hat{\Gamma}_{1101}}}\leq a, & \frac{C^3(\rho+C^3c_{\hat{e}^1\,_{01}})c_{F^\alpha}c_{\phi^\alpha_{11}}}{c_{\hat{\Gamma}_{1111}}}\leq a, & \frac{4(c_{\phi^\alpha_{11}}+C^3c_{\hat{e}^1\,_{11}}c_{\phi^\alpha_{00}})}{c_{\phi^\alpha_{00}}}\leq a,\\
  \frac{(\rho+C^3c_{\hat{e}^1\,_{01}})c_{\phi^\alpha_{00}}}{c_{\phi^\alpha_{01}}}\leq a, & \frac{(\rho+C^3c_{\hat{e}^1\,_{01}})c_{\phi^\alpha_{01}}}{c_{\phi^\alpha_{11}}}\leq a.
 \end{array}
\end{eqnarray*}

We define
\begin{equation*}
 \rho \equiv \max\{\rho_{\phi^\alpha},\rho_{\phi^\alpha_{00}}\} \leq \frac{1}{8}
\end{equation*}
and
\begin{equation*}
 a \equiv (32\rho^2)^\frac{1}{3}<1.
\end{equation*}
Considering the last two inequalities, we define
\begin{equation*}
 c_{\hat{e}^1\,_{01}}\equiv\frac{\rho}{C^3},\hspace{1cm}c_{\phi^\alpha_{01}}\equiv\frac{2\rho c_{\phi^\alpha_{00}}}{a},\hspace{1cm}c_{\phi^\alpha_{11}}\equiv\frac{2\rho c_{\phi^\alpha_{01}}}{a}.
\end{equation*}
With the previous definitions and
\begin{equation*}
 c_{\hat{e}^1\,_{11}}\equiv\frac{4\rho^2}{C^3a^2}
\end{equation*}
the third to last inequality is satisfied. The rest of the inequalities are satisfied by defining
\begin{equation*}
 \begin{array}{lll}
  c_{\hat{\Gamma}_{0100}}\equiv\frac{C^3c_{F^\alpha}c_{\phi^\alpha_{00}}}{2a}, & c_{\hat{\Gamma}_{0101}}\equiv\frac{C^3c_{F^\alpha}c_{\phi^\alpha_{01}}}{2a}, & c_{\hat{\Gamma}_{0111}}\equiv\frac{C^3c_{F^\alpha}c_{\phi^\alpha_{11}}}{2a},\\
  c_{\hat{\Gamma}_{1100}}\equiv\frac{C^3c_{F^\alpha}c_{\phi^\alpha_{01}}}{a}, & c_{\hat{\Gamma}_{1101}}\equiv\frac{C^3c_{F^\alpha}c_{\phi^\alpha_{11}}}{a}, & c_{\hat{\Gamma}_{1111}}\equiv\frac{C^3(\rho+C^3c_{\hat{e}^1\,_{01}})c_{F^\alpha}c_{\phi^\alpha_{11}}}{a},\\
  c_{\hat{e}^2\,_{01}}\equiv\frac{c_{\hat{\Gamma}_{0100}}}{2a}, & c_{\hat{e}^2\,_{11}}\equiv\frac{c_{\hat{\Gamma}_{1100}}}{a}.
 \end{array}
\end{equation*}
Now we can take $r$ big enough so that all the $A$'s are less or equal than $1$.
\end{proof}

The following lemma states the convergence result. The proof follows as the one given in \cite{Friedrich07}.
\begin{lem}\label{holomorphicSolutions}
The estimates \ref{convergenceEstimates} for the derivatives of the functions $X$ and the expansion types given in Lemma \ref{expansionTypeFields} imply that the associated Taylor series are absolutely convergent in the domain $|v|<\frac{1}{\alpha\rho}$, $|u|+|w|<\frac{\alpha^2}{r}$, for any real number $\alpha$, $0<\alpha\leq 1$. It follows that the formal expansions determined in Lemma \ref{expansionCoefficients} define indeed a (unique) holomorphic solution to the $\partial_u$-equations, the $\partial_w$-equations and the $\sigma^\alpha_{00}=0$ equations, which induces the data $\phi^\alpha$ on $W_0$.
\end{lem}

\section{The complete set of equations on $\hat{N}$}\label{completeSet}

We have seen in Section \ref{sectionConformalEquations} how to calculate a formal expansion for our fields using a subset of the field equations. In the previous section we have shown that these formal expansions are convergent in a neighbourhood of $q$. In this section we shall show that these fields satisfy the complete system of field equations. First, we prove that the field equations are satisfied in the limit as $u\rightarrow 0$. Second, we derive a subsidiary system of equations, for which the first result provides the initial conditions, and which allows us to prove that the complete system is satisfied.
\begin{lem}\label{limitCompleteSystem}
The functions $\hat{e}^a\,_{AB}$, $\hat{\Gamma}_{ABCD}$, $\phi^\alpha$, $\phi^\alpha_{AB}$, whose expansion coefficients are determined by Lemma \ref{expansionCoefficients}, with expansions that converge on an open neighbourhood of the point $O$, neighbourhood that we assume to coincide with $\hat{N}$, satisfy the complete set of field equations on the set $U_0$ in the sense that the fields $t_{AB}\,^{CD}\,_{EF}$, $R_{ABCDEF}$, $\sigma^\alpha_{AB}$, $\Sigma^\alpha_{AB}$ calculated from these functions on $\hat{N}\backslash U_0$ have vanishing limit as $u\rightarrow 0$.
\end{lem}
\begin{proof}
 Taking into account that the equations used to calculate the expansion coefficients are already satisfied, it is left to show that $t_{01}\,^{AB}\,_{11}=0$, $R_{AB0111}=0$, $\sigma^\alpha_{01}=0$ and $\Sigma^\alpha_{AB}=0$ in the limit $u\rightarrow 0$. From the definition (see \cite{Friedrich07} for details) and using the way in which the coordinates and the frame field were constructed
\begin{equation*}
 \lim_{u\to 0}t_{01}\,^{AB}\,_{11}=0.
\end{equation*}
Using the definitions from $R_{ABCDEF}$, $r_{ABCDEF}$, $t_{AB}\,^{CD}\,_{EF}$ and the way in which the coordinates and the frame field were constructed, we have near $u=0$
\begin{eqnarray*}
\hspace{-2cm} R_{AB0111} & = & \frac{1}{2u}\Bigg[\partial_v\hat{\Gamma}_{11AB}-2\hat{\Gamma}_{111(A}\epsilon_{B)}\,^0+\epsilon_A\,^0\epsilon_B\,^0\Big(-\frac{2}{u}\hat{e}^1\,_{11}-\partial_v\hat{e}^2\,_{11}+4\hat{\Gamma}_{0111}\Big)\Bigg]\\
 && -\frac{1}{2}\Bigg[F^\alpha D_{AB}\phi^\alpha_{11}+F^{\alpha\beta}\phi^\alpha_{AB}\phi^\beta_{11}+\epsilon_A\,^0\epsilon_B\,^0\Big(\tilde{f}-\frac{1}{6}\hat{f}\Big)\Bigg]+O(u).
\end{eqnarray*}
Taking the limit $u\rightarrow 0$ and using on $U_0$ the $\partial_u$-equations, the $\partial_w$-equation and the spinorial behaviour of the quantities involved, we get
\begin{equation*}
 \lim_{u\to 0}R_{AB0111}=0.
\end{equation*}
Using that $\phi^\alpha_{01}=\frac{1}{2}\partial_v\phi^\alpha_{00}$ in $U_0$ and $\phi^\alpha_{00}=\partial_u\phi^\alpha$ as $\sigma^\alpha_{00}=0$,
\begin{equation*}
 \lim_{u\to 0}\sigma^\alpha_{01}=\Big(\frac{1}{2}\partial_u\partial_v\phi^\alpha-\phi^\alpha_{01}\Big)\Big|_{u=0}=0.
\end{equation*}
Now, as $\sigma^\alpha_{AB}|_{u=0}=0$, then $\Sigma^\alpha_{AB}|_{u=0}=(-\Sigma^\alpha_{BA}+t_{AG}\,^{EFG}\,_BD_{EF}\phi^\alpha)|_{u=0}$, which imply
\begin{equation*}
 \lim_{u\to 0}\Sigma^\alpha_{AB}=0.
\end{equation*}
This finishes the proof showing that the complete system of field equations are satisfied in the limit as $u\rightarrow 0$.
\end{proof}
\begin{lem}
The functions $\hat{e}^a\,_{AB}$, $\hat{\Gamma}_{ABCD}$, $\phi^\alpha$, $\phi^\alpha_{AB}$, corresponding to the expansions determined in Lemma \ref{expansionCoefficients}, satisfy the complete set of field equations on the set $\hat{N}$.
\end{lem}
\begin{proof}
 We have seen that the field equations are satisfied in the limit $u\rightarrow 0$, we proceed to deduce a system of equations that those quantities satisfy.

Following the proof of Lemma 5.5 in \cite{Friedrich07} we find that
\begin{equation}\label{auxt}
 \Big(\partial_u+\frac{1}{u}\Big)t_{01}\,^{AB}\,_{11}=2\hat{\Gamma}_{0100}t_{01}\,^{AB}\,_{11}+2R^{(A}\,_{00111}\epsilon_0\,^{B)}.
\end{equation}
Also following the proof of Lemma 5.5 in \cite{Friedrich07} we get in our case that
\begin{eqnarray}\label{auxR}
 \Big(\partial_u+\frac{1}{u}\Big)R_{AB0111} & = & 2\hat{\Gamma}_{0100}R_{AB0111}-\Big(S_{AB00}-\frac{1}{6}R\epsilon_A\,^1\epsilon_B\,^1\Big)t_{01}\,^{01}\,_{11}\\
 \nonumber && +\frac{1}{2}\Big(D^{EF}S_{ABEF}-\frac{1}{6}D_{AB}R\Big).
\end{eqnarray}
Let us consider the coupled system of PDE's \ref{auxt} and (\ref{auxR}), with the initial conditions already derived. Taking first \ref{auxt} with $AB=11$ we see that $t_{01}\,^{11}\,_{11}=0$. Now \ref{auxt} with $AB=01$ and (\ref{auxR}) with $AB=00$ show that 
$t_{01}\,^{01}\,_{11}=0$ and $R_{000111}=0$ if and only if $D^{EF}S_{00EF}-\frac{1}{6}D_{00}R=0$. If $D^{EF}S_{01EF}-\frac{1}{6}D_{01}R=0$ equation (\ref{auxR}) with $AB=01$ implies $R_{010111}=0$, and then \ref{auxt} with $AB=00$ implies that $t_{01}\,^{00}\,_{11}=0$. Finally, if $D^{EF}S_{11EF}-\frac{1}{6}D_{11}R=0$ then (\ref{auxR}) with $AB=11$ implies $R_{110111}=0$.

In the equations $R_{AB00EF}=0$, which where used to calculate the unknowns as part of the $\partial_u$-equations, the quantities $S_{0000}$, $S_{0001}$, $S_{0011}$, $S_{0111}$ and $R$ are equal to the corresponding components of the trace-free part of the Ricci spinor and the Ricci scalar. Then the equations $D^{EF}S_{00EF}-\frac{1}{6}D_{00}R=0$ and $D^{EF}S_{01EF}-\frac{1}{6}D_{01}R=0$ are automatically satisfied, as they only include $S_{0000}$, $S_{0001}$, $S_{0011}$, $S_{0111}$ and $R$, and if these quantities are replaced for their expression in terms of the field frame coefficients and the connection coefficients, then the two equations are the components of the contracted Bianchi identity. The equation that is not automatically satisfied is $D^{EF}S_{11EF}-\frac{1}{6}D_{11}R=0$, as it includes $S_{1111}$, which has not been used as part of the procedure to calculate the unknowns. So we need to include
\begin{equation}\label{requirement}
 D^{EF}S_{11EF}-\frac{1}{6}D_{11}R=0
\end{equation}
as requirement for the complete system of field equations to be satisfied. So we have that if and only if \ref{requirement} is satisfied then
\begin{equation*}
 t_{AB}\,^{EF}\,_{CD}=0,\hspace{2cm}R_{ABCDEF}=0.
\end{equation*}
We assume from now on \ref{requirement} to be satisfied and analyze the implications in the following section.

From the definition of $\sigma^\alpha_{AB}$,
\begin{equation*}
 D_{AB}\sigma^\alpha_{CD}-D_{CD}\sigma^\alpha_{AB}=-t_{AB}\,^{EF}\,_{CD}D_{EF}\phi^\alpha+\epsilon_{BC}\Sigma^\alpha_{AD}+\epsilon_{AD}\Sigma^\alpha_{CB},
\end{equation*}
then
\begin{equation*}
 \Big(\partial_u+\frac{1}{u}\Big)\sigma^\alpha_{01}=2\hat{\Gamma}_{0100}\sigma^\alpha_{01},
\end{equation*}
which together with the initial condition for $u=0$ gives $\sigma^\alpha_{01}=0$, and together with the quantities that are already known to be zero from the $\partial_u$-equations
\begin{equation*}
 \sigma^\alpha_{AB}=0.
\end{equation*}
The last equation imply that
\begin{equation*}
 \Sigma^\alpha_{AB}=-\Sigma^\alpha_{BA}+t_{AG}\,^{EFG}\,_BD_{EF}\phi^\alpha,
\end{equation*}
 using the $\partial_u$-equations and what has already been deduced,
\begin{equation*}
 \Sigma^\alpha_{AB}=0.
\end{equation*}
So the full system of field equations is satisfied.
\end{proof}

\section{Conditions on $f$, $f^\alpha$, $F^\alpha$, $F^{\alpha\beta}$}

We need to consider now the requirement \ref{requirement}. If we write it in full using the field equations, it takes the form
\begin{equation}\label{req2}
 J^{\alpha AB}D_{11}D_{AB}\phi^\alpha+K^\alpha D_{11}\phi^\alpha=0,
\end{equation}
where $J^{\alpha AB}$ is the spinor version of
\begin{equation}\label{Jalpha}
 J^{\alpha a}=-\frac{1}{2}\frac{\partial f}{\partial(D_a\phi^\alpha)}+\frac{1}{2}F^\beta\frac{\partial f^\beta}{\partial(D_a\phi^\alpha)}+\Big(\frac{\partial F^\alpha}{\partial\phi^\beta}+F^{\alpha}F^{\beta}\Big)D^a\phi^\beta
\end{equation}
and
\begin{eqnarray}\label{Kalpha}
 K^\alpha & = & -\frac{1}{2}\frac{\partial f}{\partial\phi^\alpha}+fF^\alpha+\frac{1}{2}F^\beta\frac{\partial f^\beta}{\partial \phi^\alpha}-\frac{1}{2}f^\beta\frac{\partial F^\beta}{\partial \phi^\alpha}+f^\beta F^{\beta\alpha}\\
 \nonumber && +\Big(\frac{\partial F^{\alpha\beta}}{\partial\phi^\gamma}-\frac{1}{2}\frac{\partial F^{\beta\gamma}}{\partial\phi^\alpha}+F^{\alpha\gamma}F^\beta\Big)D_a\phi^\beta D^a\phi^\gamma.
\end{eqnarray}
As
\begin{equation*}
 D_{AB}D_{CD}\phi^\alpha=D_{(AB}D_{CD)}\phi^\alpha+\frac{1}{3}f^\alpha h_{ABCD},
\end{equation*}
we can write \ref{req2} in the following form,
\begin{equation*}
\hspace{-1.5cm} J^\alpha_{00}D_{(11}D_{11)}\phi^\alpha-2J^\alpha_{01}D_{(01}D_{11)}\phi^\alpha+J^\alpha_{11}\Big(\frac{1}{3}f^\alpha+D_{(00}D_{11)}\phi^\alpha\Big)+K^\alpha D_{11}\phi^\alpha=0.
\end{equation*}
If we now evaluate this expression at the origin all the quantities involved depend only on the free initial data. That is, $J^\alpha_{AB}$, $K^\alpha$ and $f^\alpha$ are functions of $\phi^\alpha|_0$ and $D_{AB}\phi^\alpha|_0$. As $D_{(AB}D_{CD)}\phi^\alpha|_0$ are also part of the free initial data, in order for the equality to be satisfied for all initial data we need that
\begin{equation}\label{req3}
 J^\alpha_{AB}=0.
\end{equation}
So from \ref{req2} we are left with
\begin{equation*}
 K^\alpha D_{11}\phi^\alpha=0.
\end{equation*}
Again evaluating this expression at the origin, as $K^\alpha$ is a scalar function, and considering that the equality should be satisfied for all orthonormal frames at the origin, we also need that
\begin{equation}\label{req4}
 K^\alpha=0.
\end{equation}
Conditions \ref{req3} and \ref{req4} imply that the contracted Bianchi identity needs to be satisfied by both sides of \ref{Riccih} from the beginning.

Let us consider \ref{req3}. $F^\alpha$ does not depend on $D_a\phi^\alpha$, then the condition using the expression \ref{Jalpha} can be written as
\begin{equation*}
 \frac{\partial}{\partial(D_a\phi^\alpha)}(f-F^\beta f^\beta)=2\Big(\frac{\partial F^\alpha}{\partial\phi^\beta}+F^{\alpha}F^{\beta}\Big)D^a\phi^\beta.
\end{equation*}
This equation can only be integrated if
\begin{equation}\label{intCond}
 \frac{\partial F^\alpha}{\partial\phi^\beta}=\frac{\partial F^\beta}{\partial\phi^\alpha},
\end{equation}
and then
\begin{equation*}
 f-F^\alpha f^\alpha=\Big(\frac{\partial F^\alpha}{\partial\phi^\beta}+F^{\alpha}F^{\beta}\Big)D_a\phi^\alpha D^a\phi^\beta+G(\phi^\gamma),
\end{equation*}
where $G(\phi^\gamma)$ is a free function of $\phi^\gamma$. We can consider this equation as an expression for $f$ in terms of the other functions, having
\begin{equation*}
 f=F^\alpha f^\alpha+\Big(\frac{\partial F^\alpha}{\partial\phi^\beta}+F^{\alpha}F^{\beta}\Big)D_a\phi^\alpha D^a\phi^\beta+G(\phi^\gamma).
\end{equation*}
If we put this into \ref{req4} using the expression (\ref{Kalpha}) we get
\begin{eqnarray*}
 && \frac{1}{2}\frac{\partial G}{\partial\phi^\alpha}-GF^\alpha+f^\beta\Big(\frac{\partial F^\beta}{\partial\phi^\alpha}-F^\alpha F^\beta-F^{\alpha\beta}\Big)+\Big(\frac{1}{2}\frac{\partial^2F^\gamma}{\partial\phi^\alpha\partial\phi^\beta}-F^\alpha\frac{\partial F^\beta}{\partial\phi^\gamma}\\
 && +F^\beta\frac{\partial F^\gamma}{\partial\phi^\alpha}-\frac{\partial F^{\alpha\beta}}{\partial\phi^\gamma}+\frac{1}{2}\frac{\partial F^{\beta\gamma}}{\partial\phi^\alpha}-F^\alpha F^\beta F^\gamma-F^{\alpha\beta}F^\gamma\Big)D^a\phi^\beta D_a\phi^\gamma=0.
\end{eqnarray*}
This can be written in the following form
\begin{equation}\label{req5}
 G^{\alpha\beta}f^\beta=G^\alpha+G^{\alpha\beta\gamma}D^a\phi^\beta D_a\phi^\gamma,
\end{equation}
where
\begin{eqnarray*}
 G^\alpha & = & -\frac{1}{2}\frac{\partial G}{\partial\phi^\alpha}+F^\alpha G,\\
 G^{\alpha\beta} & = & \frac{\partial F^\beta}{\partial\phi^\alpha}-F^\alpha F^\beta-F^{\alpha\beta},\\
 G^{\alpha\beta\gamma} & = & -\frac{1}{2}\frac{\partial^2F^\gamma}{\partial\phi^\alpha\partial\phi^\beta}+F^\alpha\frac{\partial F^\beta}{\partial\phi^\gamma}-F^\beta\frac{\partial F^\gamma}{\partial\phi^\alpha}\\
 && +\frac{\partial F^{\alpha\beta}}{\partial\phi^\gamma}-\frac{1}{2}\frac{\partial F^{\beta\gamma}}{\partial\phi^\alpha}+F^\alpha F^\beta F^\gamma+F^{\alpha\beta}F^\gamma.
\end{eqnarray*}
If $\det(G^{\alpha\beta})\neq 0$ then
\begin{equation*}
 f^\rho=(G^{\rho\alpha})^{-1}(G^\alpha+G^{\alpha\beta\gamma}D^a\phi^\beta D_a\phi^\gamma)
\end{equation*}
Equation \ref{intCond} means that the form corresponding to $F^\alpha$ is closed, and therefore there exists a function $F(\phi^\gamma)$ such that $F^\alpha=\frac{\partial F}{\partial\phi^\alpha}$.

So we are left with the possibility of choosing $\frac{1}{2}(n^2+n+4)$ free functions of the fields $\phi^\gamma$, namely $G$, $F$, $F^{\alpha\beta}$. Once we have these functions, we calculate $F^\alpha=\frac{\partial F}{\partial\phi^\alpha}$, and if $\det(G^{\alpha\beta})\neq 0$ we calculate $f^\alpha$. If $\det(G^{\alpha\beta})=0$ then some of the $f^\alpha$'s are free functions, and we have conditions on the r.h.s. of \ref{req5}.

\section{Analyticity at $q$}\label{analyticity}

The last needed step is to show that the holomorphic solution of Lemma \ref{holomorphicSolutions} can be extended as to cover a full neighbourhood of the point $q$. This is not obvious from our construction as our gauge is singular. The proof that we can indeed get a holomorphic solution in a hole neighbourhood of $q$ is in all similar to the respective proofs in \cite{Friedrich07,Acena09}, therefore here we only explain the steps of the proof.

The first step is to show that the obtained solution can be expressed in terms of the normal coordinates $x^a$ and the frame field $c_{AB}$ based on the frame $c_{AB}$ at $q$. This can be done by showing that for small $|x^c|$ the coordinate transformation $x^a\rightarrow z^a(x^c)$, where defined, is nondegenerate. This means that all the tensor fields entering the field equations can be expressed in terms of the normal coordinates $x^a$ and the normal frame field $c_{AB}$. The coordinates $x^a$ cover a domain $U$ in $\mathbb{C}^3$ on which the frame vector fields $c_{AB}=c^a\,_{AB}\partial_{x^a}$ exist, are linearly independent and holomorphic. Also in $U$ the other tensor fields expressed in terms of the $x^a$ and $c_{AB}$ are holomorphic. However, by the singularity of our gauge, $U$ does not contain the hypersurface $x^1+ix^2=0$ but the boundary of $U$ becomes tangent to this hypersurface at $x^a=0$.

The second step is to notice that the construction of the submanifold $\hat{N}$ was done based on the frame $c_{AB}$ at $q$. Starting with a different frame $\tilde{c}_{AB}$ at $q$ all the previous constructions and derivations can be repeated if the estimates for the null data in the $c_{AB}$-gauge can be translated to the same type of estimates for the null data in the $\tilde{c}_{AB}$-gauge. Indeed, the estimates do translate into each other, and hence all the statements made about the solution in the $c_{AB}$-gauge apply to the solution in the $\tilde{c}_{AB}$-gauge, in particular statements about domains of convergence.

Now, the solution in the $\tilde{c}_{AB}$-gauge can be expressed in terms of normal coordinates $\tilde{x}^a$ based on the frame $\tilde{c}_{AB}$ at $q$, which cover a domain $\tilde{U}$ in $\mathbb{C}^3$. Again $\tilde{U}$ does not contain the hypersurface $\tilde{x}^1+i\tilde{x}^2=0$ but the boundary of $\tilde{U}$ becomes tangent to this hypersurface at $\tilde{x}^a=0$.

By the uniqueness statements made so far, the solution in the $\tilde{c}_{AB}$-gauge and in normal coordinates $\tilde{x}^a$ is related to the solution in the $c_{AB}$-gauge and in normal coordinates $x^a$, on the domain $U\cap\tilde{U}$, by the rotation that takes $\tilde{c}_{AB}$ to $c_{AB}$, corresponding to the rotation of normal coordinates. We can extend this as a coordinate and frame transformation to the solution obtained in the $\tilde{c}_{AB}$-gauge to express all fields in terms of $x^a$ and $c_{AB}$. Then the solution obtained in the $c_{AB}$-gauge and the solution in the $\tilde{c}_{AB}$-gauge are genuine holomorphic extensions of each other, as one covers the singular generator of the other one away from the origin in a regular way.

Therefore, the set $U$ can be extended in such a way as to contain a punctured neighbourhood of the origin in which the solution is holomorphic in the normal coordinates $x^a$ and the normal frame $c_{AB}$. Then the solution is in fact holomorphic on a full neighbourhood of the origin $x^a=0$, which represents the point $q$, as holomorphic functions in more than one dimension cannot have isolated singularities.

By Lemma \ref{formalExpansion} we have from null data satisfying the reality conditions a formal expansion of the solution with expansion coefficients satisfying the reality conditions. By the various uniqueness statements obtained in the lemmas, this expansion must coincide with the expansion in normal coordinates of the solution obtained above. This implies the existence of a 3-dimensional real slice on which the tensor fields satisfy the reality conditions. It is obtained by requiring the coordinates $x^a$ to assume values in $\mathbb{R}^3$. This completes the proof of Theorem \ref{mainThm}.

\section{Conclusions}

We have seen how to determine a formal expansion of the solution to certain types of elliptic systems of equations at  a given point using a minimal set of freely specifiable data, the null data. We have also obtained necessary and sufficient conditions on the null data for the formal expansion to be absolutely convergent, thus showing that the null data characterize all possible solutions in a neighbourhood of the given point.

The system of equations is general enough as to include as particular cases stationary Einstein-Maxwell fields, static Einstein-Maxwell-dilaton fields and harmonic maps coupled to gravity whose base space is three dimensional or where the spacetime is stationary.

One interesting outcome of the analysis are the conditions that the functions of the fields that enter the field equations need to satisfy for the existence of solutions. These conditions can be read as that the contracted Bianchi identity needs to be fulfilled by the system of equations. This seems to indicate that only geometrically well behaved systems of equations allow for general data and solutions. It could be interesting to analyze if this forces the system of equations to be the Euler-Lagrange equations for some Lagrangian. Along the same line, it would also be interesting to analyze whether the considered system of equations can generally arise from a dimensional reduction. This would be the case for example if one starts with a four dimensional stationary spacetime or in Kaluza-Klein type theories.

\section*{Acknowledgements}

I would like to thank Helmut Friedrich for helpful discussions.

\section*{References}
\bibliographystyle{abbrv}
\bibliography{bib_minimal_data_geometric_systems_Acena}

\end{document}